\documentclass[]{IEEEtran}
\usepackage{graphicx}

\usepackage{latexsym}

\usepackage{hyperref}

\usepackage{xcolor}
\definecolor{mpsNavy}{rgb}{0.01,0.01,0.5}
\definecolor{mpsGray}{rgb}{0.85,0.84,0.84}
\definecolor{mpsMyrtleGreen}{rgb}{0.19,0.47,0.45}
\definecolor{huiGray}{rgb}{0.97,0.97,0.97}
\definecolor{chromeyellow}{rgb}{1.0, 0.65, 0.0}
\definecolor{VG}{rgb}{0.6,0.6,0.6}

\usepackage{subfigure}

\usepackage[numbers, square]{natbib}

\usepackage{booktabs}
\usepackage{enumitem}
\usepackage{tabu}

\usepackage[np]{numprint}
\npthousandsep{,}
\npdecimalsign{.}

\definecolor{cerulean}{rgb}{0.105, 0.672, 0.836}

\usepackage{soul}

\usepackage{paralist}

\usepackage{xspace}
\newcommand{\etal}{et al.\xspace}

\newcommand{\fsc}{\textsc}
\newcommand{\fsl}{\textsl}

\newcommand{\fsub}[1]{\texorpdfstring{\textsubscript{#1}}{#1}}

\newcommand{\approach}{\fsc{MissAuditor}\xspace}

\newcounter{textboxno}
\usepackage[many]{tcolorbox}

\newtcolorbox{mybox}[2][parbox=false]{%
boxsep=3pt,left=2pt,right=2pt,bottom=5pt,
width=\columnwidth,
boxrule=1pt,
attach boxed title to top center = {yshift=-\tcboxedtitleheight/2},
colbacktitle=white,coltitle=black,
boxed title style={size=normal,colframe=white,boxrule=0pt}, 
interior style={white},
title={\refstepcounter{textboxno}\label{#1}
Example \arabic{textboxno}: {#2}
\def\@currentlabel{\p@textboxno\thetextboxno}},
enhanced,
float,
}

\begin{document}
\newcolumntype{P}[1]{>{\centering\arraybackslash}p{#1}}
\title{Understanding Mobile App Reviews to Guide Misuse Audits}

\author{
% Vaibhav Garg, 
% Hui Guo, 
% Nirav Ajmeri, 
% Saikath Bhattarcharya, 
% Munindar P. Singh \\
% vaibhavg@vt.edu,
% hguo@quora.com,
% nirav.ajmeri.bristol.ac.uk,
% bhattacharya@msoe.edu,
% mpsingh@ncsu.edu
    % \IEEEauthorblockN{Vaibhav Garg}
    % \IEEEauthorblockA{Virginia Tech\\
    % Alexandria, USA\\
    % Email: vaibhavg@vt.edu}
    % \and
    % \IEEEauthorblockN{Hui Guo}
    % \IEEEauthorblockA{Quora Inc.\\
    % Mountain View, California, USA\\
    % Email: ncsuguo@gmail.com}
    % \and
    % \IEEEauthorblockN{Nirav Ajmeri}
    % \IEEEauthorblockA{University of Bristol\\
    % Bristol, UK\\
    % Email: nirav.ajmeri@bristol.ac.uk}
    % \and
    % \IEEEauthorblockN{Saikath Bhattacharya}
    % \IEEEauthorblockA{Milwaukee School of Engineering\\
    % Milwaukee, Wisconsin, USA\\
    % Email: bhattacharya@msoe.edu}
    % \and
    % \IEEEauthorblockN{Munindar P. Singh}
    % \IEEEauthorblockA{North Carolina State University\\
    % Raleigh, North Carolina, USA\\
    % Email: mpsingh@ncsu.edu}
    \IEEEauthorblockN{Vaibhav Garg\IEEEauthorrefmark{1}, Hui Guo\IEEEauthorrefmark{2}, Nirav Ajmeri\IEEEauthorrefmark{3}, Saikath Bhattacharya\IEEEauthorrefmark{4}, Munindar P. Singh\IEEEauthorrefmark{5}}
    
    \IEEEauthorblockA{\IEEEauthorrefmark{1}Virginia Tech, Alexandria, USA\\
    Email: vaibhavg@vt.edu}
    
    \IEEEauthorblockA{\IEEEauthorrefmark{2}Quora Inc., Mountain View, California, USA\\
    Email: ncsuguo@gmail.com}
    
    \IEEEauthorblockA{\IEEEauthorrefmark{3}University of Bristol, Bristol, UK\\
    Email: nirav.ajmeri@bristol.ac.uk}
    
    \IEEEauthorblockA{\IEEEauthorrefmark{4}Milwaukee School of Engineering, Milwaukee, Wisconsin, USA\\
    Email: bhattacharya@msoe.edu}
    
    \IEEEauthorblockA{\IEEEauthorrefmark{5}North Carolina State University, Raleigh, North Carolina, USA\\
    Email: mpsingh@ncsu.edu}    
}

\maketitle
\pagestyle{plain}
\thispagestyle{plain}

\begin{abstract}

\textbf{Problem:} 
We address the challenge in responsible computing where an \emph{exploitable} mobile app is misused by one app user (an \emph{abuser}) against another user or bystander (\emph{victim}). 
We introduce the idea of a \emph{misuse audit} of apps as a way of determining if they are exploitable without access to their implementation. 
% \textbf{Background:} 
% Misuse by app users (as opposed to app developers) falls out of the scope of current methods such as software audits, which in any case apply only when the implementation of an app is available.  
\textbf{Method:} 
We leverage app reviews to identify exploitable apps and their functionalities that enable misuse. 
First, we build a computational model to identify alarming reviews (which report misuse). Second, using the model, we identify exploitable apps and their functionalities. Third, we validate them through manual inspection of reviews. 
\textbf{Findings:} 
Stories by abusers and victims mostly focus on past misuses, whereas stories by third parties mostly identify stories indicating the potential for misuse. 
Surprisingly, positive reviews by abusers, which exhibit language with high dominance, also reveal misuses. In total, we confirmed \np{156} exploitable apps facilitating the misuse. Based on our qualitative analysis, we found exploitable apps exhibiting four types of exploitable functionalities. 
\textbf{Implications:} Our method can help identify exploitable apps and their functionalities, facilitating misuse audits of a large pool of apps. 
\end{abstract}


%    <concept>
%        <concept_id>10003120.10003130.10011762</concept_id>
%        <concept_desc>Human-centered computing~Empirical studies in collaborative and social computing</concept_desc>
%        <concept_significance>500</concept_significance>
%        </concept>
%    </ccs2012>
%    <concept>
%        <concept_id>10003456.10003462</concept_id>
%        <concept_desc>Social and professional topics~Computing / technology policy</concept_desc>
%        <concept_significance>300</concept_significance>
%        </concept>

% \ccsdesc[300]{Social and professional topics~Computing / technology policy}
% \ccsdesc[500]{Human-centered computing~Empirical studies in collaborative and social computing}

% \keywords{mobile apps, auditing, app reviews}

\section{Introduction}
\label{sec:intro}
%%%%%%%%%%%%%%%%%%%%%%%%%

Traditional audits of mobile apps conduct a review of their source code \citep{Bluejay2021, EffectiveAudits2019}. However, interpersonal misuse arising from app users (instead of app developers) goes unnoticed by such processes. We introduce \emph{misuse audit}, a process for auditing mobile apps through the reported misuse cases. 

A misuse occurs when an app user (\emph{abuser}) exploits the app to access information of other users or third parties (\emph{victims}). In particular, a misuse audit can identify cases when the victim (1) doesn't know about the information access (spying) or (2) may know about the access but is uncomfortable with it. The latter case (often missed by traditional audits) includes incidents of forced consent or when public information (such as on dating apps) is accessed beyond the victim's level of comfort. We use the term \emph{exploitable behavior} to denote these two types of information access and use \emph{exploitable apps} to refer to apps that enable exploitable behavior. 

Research \citep{IPVspyware2018, Phonehacked2019, Clinicalsecurity2019} shows that exploitable apps may cause discomfort, fear, and potential harm to the victim.  Possible ways to prevent this risk include highlighting these apps and their exploitable functionalities to users, app distribution platforms, and app developers. However, identifying these apps is nontrivial since they have a legitimate purpose but are misused by abusers.  
We propose an approach, \approach, for misuse audits that identify exploitable apps and uncover their exploitable functionalities. 

We find that the reviews of an app often describe exploitable functionalities, its misuse (potential and actual), and users' expectations. Such reviews are evidence of exploitable behavior and can guide audits of such apps. Moreover, app reviews are more valuable than app metadata (such as app descriptions) because metadata indicates only their legitimate purpose and not the actual misuse. In particular, we propose the following research questions:

\begin{description}%[leftmargin=1em]
\item[RQ\fsub{info}.] What information regarding misuse audit is contained in reviews?
\item[RQ\fsub{identify}.] How can we conduct a misuse audit through app reviews?
\item[RQ\fsub{functionality}.] What exploitable functionalities are present in audited apps? 
\end{description}

Example~\ref{box:exploitable-behavior-relevant} shows three reviews (edited for grammar), taken from Apple's App Store \cite{Appstore} that are relevant to exploitable behaviors. Although our study is based on Apple App Store's reviews, \approach can be applied to reviews from other sources, including Google's Play Store \cite{Playstore}.

\begin{mybox}[box:exploitable-behavior-relevant]{Cases relevant to exploitable behavior}

\noindent\textbf{Fly on the wall!}\\
(for the AirBeam Video Surveillance app \cite{AirBeam})\\
\fsl{``with this app, i can spy on my family without them knowing it! it's such an awesome app!''}\bigskip

\noindent\textbf{This app basically ruined my family to an extent}\\
(for the Life360 app \cite{Life360})\\
\fsl{``My mother made everyone in the family get this app. She freaks out when the app doesn't do its job because of random obstacles that mess with the location accuracy. \ul{Drains the battery and makes my parents paranoid to know where I am at all times. I don't even do any bad stuff, yet years of trust building are being swept away by the ability to spy on the children of a household.} If you're a parent I highly recommend you don't get this app because it is extremely uncomfortable to have and it makes parents trust their children less.''}\bigskip

\noindent\textbf{Honest}\\
(for the 3Fun: Threesome \& Swingers app \cite{Threefun})\\
\fsl{``\ldots A lot of the local people I’ve talked to
(Male half of a couple) have been guys who are saying they’re part of a couple, and \ul{in all reality are single guys just looking to collect pictures}. There is no way to report that that is why you are reporting them. It’s just a boilerplate report feature. I feel there should be a way for the 3Fun community to point out people for bad behavior like this.''}
\end{mybox}

In Example~\ref{box:exploitable-behavior-relevant}, the first review, for AirBeam Video \cite{AirBeam}, addresses the scenario where the app assists a user in accessing a victim's information without the victim's knowledge. AirBeam Video is a surveillance app to be installed on the abuser's device. 
Hence, the victim may not be an app user but a bystander. 
The second review, for Life360 \cite{Life360}, complains about the problem of inappropriate access to the user's location by the user's mother. 
Due to the unequal power dynamics between the victim (reviewer in this case) and the abuser (mother in this case), the victim is forced to install apps that violate their privacy. 
The third review, from 3Fun \cite{Threefun}, describes a story of improper access to profile pictures.  
Even though the profile pictures are public, the victim is uncomfortable with the access. 
It is common for users to upload such information (pictures in this case) on an app with expectations of how other users would access it. 
As shown in these three cases, information access may lead to discomfort, fear, or potential harm \citep{Phonehacked2019,Clinicalsecurity2019}. Thus, such cases of information access indicate misuse.

\subsection{Key Findings}
We find that reviews contain rich information about apps, which is crucial in auditing them for their misuse potential. For example, not only negative but positive reviews also reveal an app's exploitable functionalities. Relevant reviews are written by abusers, victims, and third person, each type linguistically distinct, which makes mining reviews challenging (Section~\ref{sec:motivation}). Moreover, while reporting exploitability, reviews have varying degrees of \fsl{convincingness} and \fsl{severity}, which can be leveraged in misuse audit (Section~\ref{sec:approach}). We found that exploitable apps exhibit a variety of exploitable functionalities, ranging from tracking location to monitoring phone activities such as accessing chats, phone contacts, call history, and so on (Section~\ref{sec:exploitable-capabilities}). 

\subsection{Contributions and Novelty}

Our work's novelty lies in introducing the problem of misuse audits. We leverage app reviews, which are a large unexplored resource for misuse audits. We contribute to responsible computing in the following manner.

First, we show that app reviews are a viable source of information on its exploitable functionalities. Second, \approach, an approach based on app reviews for \textsc{misuse audits} of apps.
Third, a list of exploitable apps along with their alarming reviews revealing exploitable functionalities.

\subsection{Organization}
The rest of this paper is organized as follows. 
Section~\ref{sec:motivation} describes our preliminary investigation that shows that app reviews contain evidence useful for audits. 
Section~\ref{sec:approach} describes how to identify exploitable apps using our misuse audit approach, \approach. Section~\ref{sec:exploitable-capabilities} shows the procedure to uncover exploitable functionalities of audited apps. Section~\ref{sec:background} lists related work on app reviews and mobile apps. Section~\ref{sec:conclusion} concludes the paper and Section~\ref{sec:reproducibility} discusses the reproducibility of our findings.
%%%%%%%%%%%%%%%%%%%%%%%%%%%%%%%%%%%%%%%%%%%%%%%%%

\section{RQ\fsub{info}: App Reviews for Misuse Audits}
\label{sec:motivation}

We describe how we collected and investigated reviews for misuse audit, which led to several qualitative findings.

\subsection{Seed Dataset}
\label{sec:IPSdataset}

Chatterjee \etal \cite{IPVspyware2018} identified \np{2707} iOS apps as candidates for Intimate Partner Surveillance (IPS), i.e., apps used by a person to spy on their intimate partner. IPS apps are specific to intimate partners and, hence, a subclass of our concept of exploitable apps. We started our analysis from Chatterjee \etal's IPS candidate list. 

During our data collection, \np{1687} of these \np{2707} apps received at least one review on the Apple App Store, leading to a \emph{seed dataset} containing \np{11.57} million reviews. Out of \np{1687} apps in the seed dataset, only \np{210} were confirmed IPS by Chatterjee \etal.

\subsection{Investigating Reviews}
\label{sec:preliminvestigation}

Since the seed dataset contains \np{11.57} million reviews, it is impractical to manually check each review. To find the reviews revealing exploitable behavior, we sampled them using three keywords (\fsl{spy}, \fsl{stalk}, and \fsl{stealth}). 
From the \np{5287} reviews that contain at least one of our keywords, we randomly sampled \np{1000} reviews for manual scrutiny. We obtained a simple random sample to investigate if arbitrary reviews reveal any misuse cases. Moreover, we analyzed what information regarding misuse they contain. This sample is diverse: it involves \np{179} apps with between 1 and 237 reviews each.

Out of \np{1000} reviews, we found \np{403} reviews reporting exploitable behavior and \np{597} others. We found that reviews reporting exploitable behavior show variation across two dimensions: \fsl{story} and \fsl{reviewer}. Based on the story, we found reviews of the following two types:

\begin{description}%[leftmargin=1em]

\item[\emph{exploitable act}:] Reviews describe someone performing an exploitable behavior. In such reviews, the reviewer is sure about the app's exploitable functionality. For example,

\fsl{``This app is useless and it just helps overprotected parents spy on their sad kids''}

\item[\emph{potential}:] Reviews express the possibility of exploitable behavior. The reviewer may not be sure of the exploitable functionality but identifies risks with the app that can be exploitable in the future. For example,

\fsl{``\ldots Hate to be a hater. May work well to spy on the kids by `accidentally' leaving iPhone in secret place.''}

\end{description}

The functionalities actually or could be misused by abusers (in exploitable act or potential) are called \emph{exploitable functionalities}. 
The above two reviews are negative. However, we also found positive reviews indicating an exploitable app. Example~\ref{box:reviewer-identity} shows positive reviews by \emph{abusers}, who brag about the exploitable functionalities (history tracking in this case) and sometimes express their delight in the misuse. Other reviews are written by \fsl{victims}, who state their concerns and grievances (frustration at the loss of privacy), and some by \fsl{third persons} reporting the misuse of the app.

\begin{mybox}[box:reviewer-identity]{Types of reviewers reporting exploitable behavior}

\noindent\textbf{Abuser}\\
\fsl{``So much better than \emph{<other app>}. I love the history feature. My kids say it's creepy and I'm being a stalker. I don't disagree but it is really nice and convenient to be able to keep track of my kids.''}
\bigskip

\noindent\textbf{Victim}\\
\fsl{``I hate this app so much! My mother is always questioning me and if I delete it she will ground me \ldots No one wants their parents to stalk them!!''}
\bigskip

\noindent\textbf{Third Person}\\
\fsl{``\ldots I don't feel like parents should track their kids AT ALL. everyone needs a little something called trust and if you don't have it then your kids will act out and have to become sneaky \ldots I do have this app but only with my friends and we don't stalk each other \ldots''}

\end{mybox}

\begin{figure}[!htb]
\includegraphics[width=0.9\columnwidth]{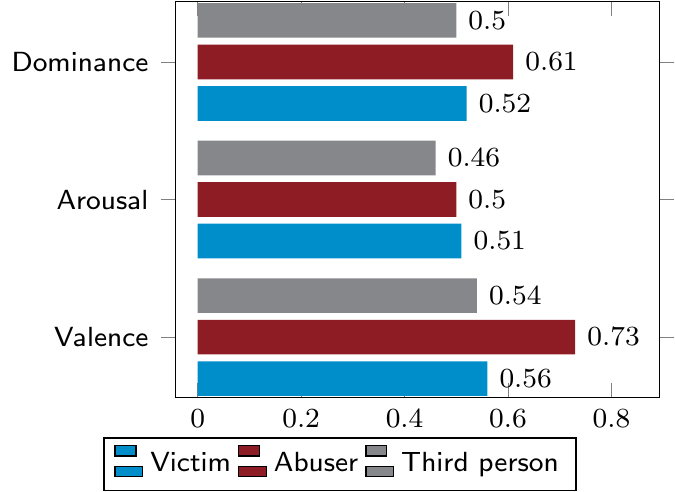}
\caption{Affective analysis showing valence, arousal, and dominance scores for each type of reviewer. Abusers' language shows higher dominance and valence than the language used by victims and third persons. Mining app reviews is challenging due to such linguistic variations. 
}
\label{fig:VAD}
% \Description{A bar chart comparing valence, arousal, and dominance scores based on the language used by each reviewer type. Abusers show the highest dominance and valence, and third persons the least, with victims slightly above third persons. Victims have the highest arousal, with abusers slightly below, and third persons the least. }
\end{figure}

Our linguistic analysis reveals that reviews by abusers not only are positive (higher valence than the other two categories) but also illustrate the abuser's dominance over other parties. Figure~\ref{fig:VAD} shows this difference in the valence and dominance scores between all three reviewers. In general, these relevant reviews show linguistic variations, complicating the task.

\begin{figure}[!htb]
\noindent\includegraphics[width=0.9\columnwidth]{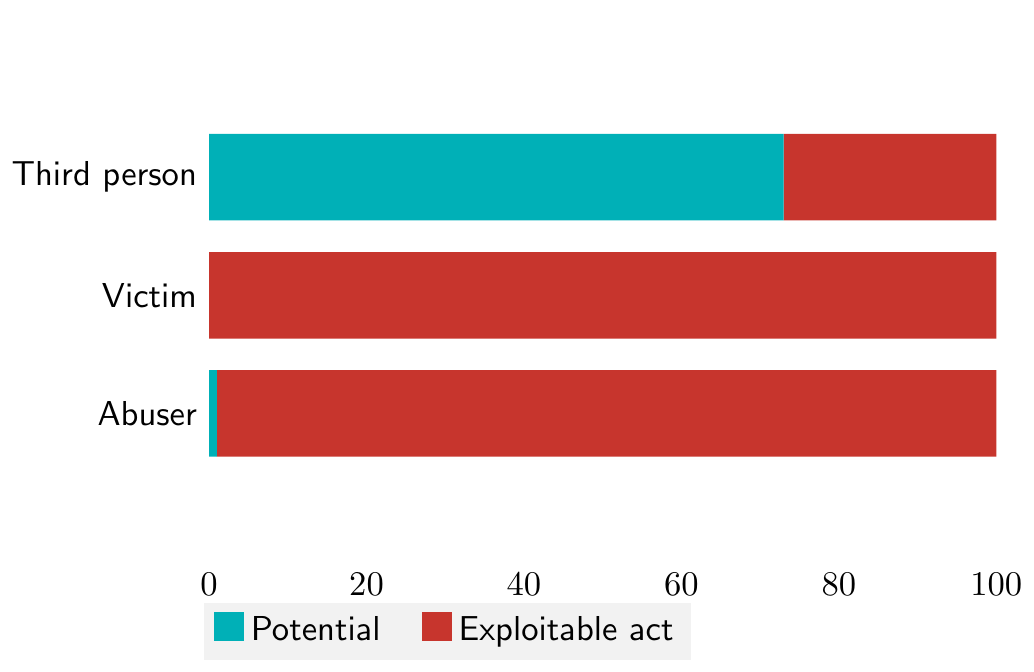}
\caption{Out of \np{403} stories manually identified as relevant, abusers and victims mostly write stories indicating exploitable acts ($\sim$\np{99}\% and \np{100}\% respectively), whereas third persons write stories indicating potential cases ($\sim$\np{73}\%).}
\label{fig:distribution}
% \Description{A stacked bar chart comparing the number of stories (exploitable acts and potential) for each reviewer type. Most of the time, abusers and victims describe exploitable acts in their reviews, whereas third-person reviewers tend to write stories indicating potential misuse.}
\end{figure}

Figure~\ref{fig:distribution} shows the relative distribution of \np{403} relevant reviews across the type of story and reviewer. The third person mostly writes potential stories (73\%), whereas abusers and victims mainly describe exploitable acts ($\sim$99\% and 100\%, respectively).

The above examples indicate that reviews contain rich information about an app's exploitable behavior. We found two types of stories that are relevant for our purposes. Moreover, such stories are written by three types of reviewers, whose language varies, making the problem of mining these reviews nontrivial. No new types of stories and reviewers were found during our extensive verification of reviews (discussed in Section~\ref{sec:conclusion}). This indicates that our random sample (\np{1000} reviews used for this analysis) was a representative set.

%%%%%%%%%%%%%%%%%%%%%%%%%%%%%%%%%%%%%%%%%%%%%%%%%
\section{RQ\fsub{identify}: Identifying Exploitable Apps through Misuse Audit}
\label{sec:approach}
\begin{figure*}[!htb]
\noindent\includegraphics[width=2\columnwidth]{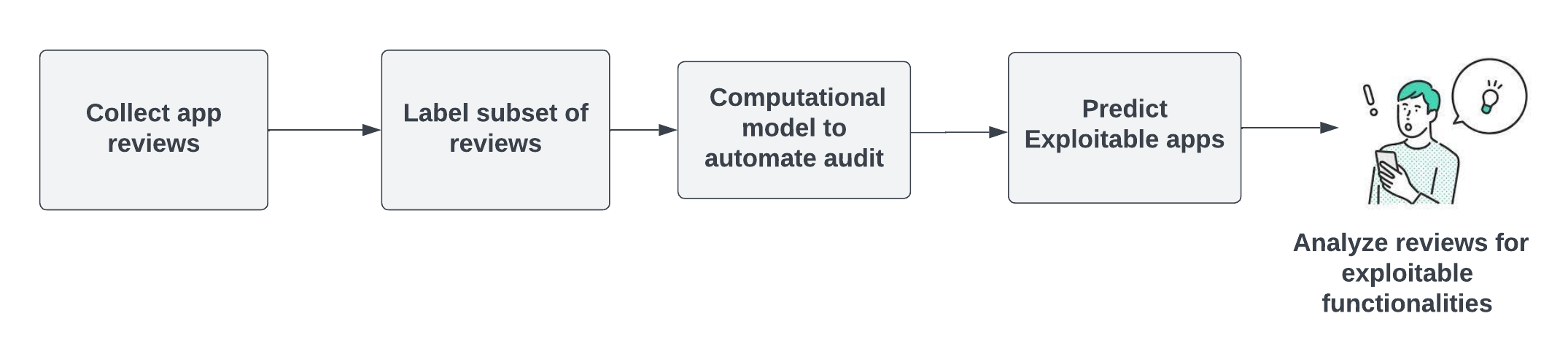}
\caption{Overview of \approach.}
\label{fig:flowdiagram}
% \Description{A series of boxes with arrows from one to the next to indicate a process pipeline. We collect app reviews from the Apple App Store. We label a small subset of the reviews for alarmingness and train a model. We apply our model on all reviews to identify exploitable apps. We manually investigate reviews of the identified apps to find their exploitable functionalities.}
\end{figure*}

We propose \approach, a review-based approach for misuse audits. Figure~\ref{fig:flowdiagram} shows an overview of the \approach approach. First, we collect reviews from the Apple App Store, as shown for the seed dataset in Section~\ref{sec:IPSdataset}. Second, we label a subset of the collected reviews for \emph{alarmingness} (defined below) and train a computational model (Section~\ref{sec:alarming}). Third, we apply our model to all reviews to identify exploitable apps (Section~\ref{sec:exploitable-score}). Fourth, we manually examine reviews of some of the identified exploitable apps to find their exploitable functionalities (Section~\ref{sec:exploitable-capabilities}).

We envision \approach to be incrementally updated by adding reviews of newly found exploitable apps. Moreover, apps with no reviews can be audited as soon as their reviews arrive.

\subsection{Alarmingness of Reviews}
\label{sec:alarming}

We find that reviews exhibit two characteristics: (i) \emph{convincingness} their claims about misuse and (ii) \emph{severity} or the effect of the misuse. We describe each of these characteristics below.

\textbf{Convincingness} depends on the review's claim about the app's exploitability. 
Some reviews report detailed exploitable behavior and, therefore, are convincing, whereas others are merely suspicions. In Example~\ref{box:convincingexample}, the first review is unrelated to exploitable behavior and, hence, is not convincing. The second review describes the reviewer's suspicion of the app, which may or may not be true (slightly convincing). The third review (by an abuser) confirms the exploitable behavior but lacks details of the exploitable functionalities or victims. In contrast, the other reviews in Example~\ref{box:convincingexample} are convincing because they confirm exploitable behavior and mention the location feature, how to set up devices, or the victims being stalked. Extremely convincing reviews also include cases when the app is used for positive purposes, such as tracking family members or pets for safety, but shows the potential to be misused in the future. The reviews that are slightly, moderately, or extremely convincing are relevant to misuse audits. Assigning a convincingness score helps rank all reviews according to the strength of their claims.

\textbf{Severity} measures the effect of exploitable behavior on the victim. Example~\ref{box:severityexamples} shows the range of reviews varying in severity. The first review is unrelated to exploitable behavior. Thus, it is not severe. The second review shows that the exploitable act is performed with consent, making this review a slightly severe case. The third review is written by the abuser and lacks the victim's perspective to analyze the exploitable effect. We assume such acts are performed without consent and consider them moderately severe. The fourth review describes the victim's misery. The victim even says, ``This app has truthfully ruined my teenage years'' in the review, which gives solid evidence to be an extremely severe case. Moreover, in the fifth review, the victim complains that others can see when he was last active (also known as last-seen information). This is public information on each profile, but still, the victim is uncomfortable with the access. App developers should be aware of such misuse.

\begin{mybox}[box:convincingexample]{Varying degree of convincingness}
% \label{box:convincingexample}
\noindent\textbf{1: Not convincing}\\
\fsl{``It is such a great game, love it so much!''}\bigskip

\noindent\textbf{2: Slightly convincing}\\
\fsl{``Setup was a breeze. Quicktime 7 pro found it easily. Unfortunately, resolution seems much, much, lower than hoped.  Video size can not be adjusted live. Hate to be a hater. \ul{May work well to spy on the kids by `accidentally' leaving iPhone in secret place}.''}\bigskip

\noindent\textbf{3: Moderately convincing}\\
\fsl{``This app is perfect for stalking people\ldots''}\bigskip

\noindent\textbf{4: Extremely convincing}\\
\fsl{``\ul{This app is awesome for our family to keep track of where everyone is at all times!} (You can turn the location off too in case you want to be in stealth mode when buying Christmas presents too.) \ldots Even our dog knows that the alert sound when a family member arrives home means \ldots''}\bigskip

\fsl{``\ldots \ul{I use it to spy on my dogs while I'm at work}; so I use it for fun, nothing fancy. My iPad is my camera, and my iPhone is my viewer. \ldots''}\bigskip

\fsl{``\ul{bro this app is high key creepy. when i'm with my dad on his days my mom even mentions how she knew everything} i was doing and it even made my dad creeped out. if you need this app then ngl yo wack. i don't want my mom stalking me.''}

\end{mybox}

\begin{mybox}[box:severityexamples]{Varying degree of severity}

\noindent\textbf{1: Not severe}\\
\fsl{``Love the graphics so far it is a great game''}\bigskip

\noindent\textbf{2: Slightly severe}\\
\fsl{``I love this app, just great because you can time your day accordingly, I like my girlfriend knowing where I am and I love stalking her, we have fun with it.\ldots''}\bigskip

\noindent\textbf{3: Moderately severe}\\
\fsl{``This app is perfect for stalking people\ldots''}\bigskip

\noindent\textbf{4: Extremely severe}\\
\fsl{``honestly if you want your kid to rebel against you even more, this is the app for you! This app has truthfully ruined my teenage years all because my mother now has a way of tracking me down 24/7. I couldn't do the normal teenage things because I was being stalked all day \ldots''}\bigskip

\fsl{``\ldots i want to share my last seen just to my family and my girlfriend not others. please add new feature in privacy that i can share my last seen to no body except my family and girl friend 
thanks soo much !''}
\end{mybox}

To capture the above two characteristics, we define the \emph{alarmingness} of a review as the geometric mean of its convincingness and severity. An app can receive a large number of reviews. Unlike binary classification of reviews, the alarmingness score not only identifies relevant reviews for audit but also ranks them based on the likelihood and danger of the misuse. For an app, the higher the alarmingness of the review, the more useful it is in auditing and identifying misuse.

We created a training set of reviews as follows. From the seed dataset, we randomly selected two sets of about \np{1000} reviews each: those that match at least one of our keywords and those that do not. 
After removing duplicate reviews, we were left with \np{952} and \np{932} reviews, respectively, combined into our training corpus of \np{1884} reviews. Including both matching and nonmatching reviews in about equal numbers makes our training corpus unbiased toward the keywords.

From these \np{1884} reviews, we excluded the reviewers' identifiers, such as usernames. The task was to rate these reviews for convincingness and severity on a four-point Likert scale (1: not, 2: slightly, 3: moderately, 4: extremely). Two authors were annotators. For reviews rated by both, we computed the average convincingness and severity scores. This annotation study possessed minimal risk and was exempted by the Institutional Review Board (IRB) of our university. We used these annotated reviews as our training data. 

We extracted linguistic features of \np{1884} reviews through the Universal Sentence Encoder (USE) \cite{UniversalSentenceEncoder2018}, a widely used approach that has proved effective for app reviews \citep{Casper2020, USEMotivation2021}. 
Using USE features, we trained various multitarget (the targets here being convincingness and severity scores) regression models \cite{Mutioutputregression2015} and found that the Support Vector Regressor (SVR) outperforms others. We considered using the reviews' metadata, such as ratings and titles, but found the metadata to be unhelpful for identifying misuse and left it out. Not only negative ratings and titles but also reviews with positive ratings and titles indicate misuse (partly discussed in Section~\ref{sec:preliminvestigation}). For example, an abuser writes a review with the title ``Great app'' and provides five-star ratings as the app facilitates misuse. As these nondiscriminatory attributes may confuse the model, we decided not to include them while training. We used only the text of reviews for training. 

We leveraged trained SVR to predict convincingness and severity scores of all \np{11.57} million reviews in the seed dataset. The alarmingness of each review is calculated by taking the geometric mean of its predicted convincingness and severity. 

\subsection{Identifying Exploitable Apps}
\label{sec:exploitable-score}

Using statistical methods, we aggregated the alarmingness of rexviews and ranked all apps according to their reported misuse. From the seed dataset, our model predicted a total of \np{100} exploitable apps (including false positives). Moreover, our approach is not dependent on the choice of candidate apps and could be applied to any set of apps. To audit additional apps, we applied our model on (i) a dataset of similar apps and (ii) a dataset of \np{100} popular apps in the utility category. We found not only IPS but also many general-purpose exploitable apps exhibiting multiple exploitable functionalities, which are described below. 

\subsubsection{Similar Apps}
\label{sec:similarapps}
For each app determined to be exploitable from the seed dataset, we obtained recommendations for similar apps from the Apple App Store's ``You May Also Like'' feature. Through this process, we obtained \np{788} similar apps. Our motivation in using Apple-recommended apps is that these apps should offer functionalities similar to the apps classified as exploitable. Further, we collected reviews (over the period August 2008 to August 2022) of these \np{788} apps---we term this the \emph{snowball dataset}.

Our model predicts \np{90} apps as exploitable (including false positives) from the snowball dataset. Our evaluation using the snowball dataset reveals that the model yields a recall of 71.60\%, which is much higher than the other baseline approaches. In this scenario, a false negative can lead to harm through misuse, whereas a false positive just wastes effort in an unnecessary audit. Hence, high recall is more valuable than high precision.

\subsubsection{100 Popular Utility Apps}
Surveillance apps that can be misused for spying fall under the ``Utilities'' category, making it an important category to audit. We considered \np{100} popular utility apps (mentioned on the Apple's App Store page \cite{Utilitiesappstore}) and collected their reviews. Since a popular app can have a lot of reviews (over the years), collecting them can be computationally expensive. 

From these \np{100} apps, our approach, \approach, classifies only one app as exploitable. Upon examining its reviews, we found that app to be nonexploitable. We also scrutinized some apps classified as nonexploitable and found their predictions true (based on manual verification of reviews). 

Some app categories, such as payment apps, calculators, and so on, are unlikely to be misused. Hence, they don't form good candidates for identification. Iteratively auditing apps (through \approach) similar to the already identified exploitable apps (as shown in Section~\ref{sec:similarapps}) is a feasible solution to uncover a large exploitable landscape.

\subsubsection{Relevance of Findings}
\label{sec:relevance}
Some reviews in our datasets are old. For example, the seed dataset was collected over the period of July 2008 to January 2020. To check if our findings are still relevant, we randomly sampled \np{50} confirmed exploitable apps from the union of the seed and the snowball datasets. We collected new reviews (from January 2020 to April 2024) for these apps and applied our \approach approach to these reviews. Then, we scrutinized their descriptions (to know their basic functionalities)  and the alarming reviews (to know misuse cases if any), following the same process as before. We found that the new reviews of three of these \np{50} apps don't report misuse. On the other hand, the other \np{47} apps still possess exploitable functionalities causing potential or actual misuse. A high success rate (94\%; 47 of 50) on a random sample implies that most of the identified apps still facilitate misuse.

\section{RQ\fsub{functionality}: Uncovering exploitable Functionalities}
\label{sec:exploitable-capabilities}
We illustrate the idea of uncovering exploitable functionalities using the exploitable apps found from the seed dataset. First, we analyzed an app's description to understand its functionalities. Second, we analyzed the top alarming reviews (discovered by \approach) for identifying existing misuse and exploitable functionalities. Through this process, we discovered the following types of exploitable functionalities.

\begin{description}
\item[Monitoring phone activities.] Some apps monitor a victim's phone activities, such as browsing history and text messages. Such apps are installed on the victim's device and activities can be monitored on another synced device. For example, SaferKid Text Monitoring App \cite{Saferkid} allows synced devices to monitor call history, web history, texts, and so on. 

\item[Audio or video surveillance.] Some apps enable audio or video surveillance without the victim's knowledge. These apps listen, view, or record a victim's voice or actions and some of them need not be installed on the victim's phone. For example, Find My Kids: Parental control \cite{Findkids} is misused to record private conversations between people without them knowing.

\item[Tracking location.] Some Global Positioning System (GPS) apps enable tracking a victim's phone, making them uncomfortable with access. For example, Find My iPhone \cite{Findmyiphone} is a legitimate app but can be misused to spy on the location of connected devices. 

\item[Profile stalking.] Some apps are misused for stalking user profiles or user-generated content (such as text and images). For example, Kik Messaging and Chat App \cite{Kik} can be misused for stalking images and victims' other information on the app.
\end{description}

Table~\ref{tab:exploitable-capability-types} shows these four types of exploitable functionalities and some alarming reviews reporting them.
Some of these reviews are old (2014 or 2012), but we confirmed that similar concerns are being raised in the recent reviews of the same apps. For example, the Find My iPhone \cite{Findmyiphone} app lets its users see the location of the connected devices. Due to unequal power dynamics between the abuser and the victim (say in a family setting) \cite{IPVspyware2018}, the victim is sometimes forced to connect to such apps and allow their device to be located. In other words, legitimate apps designed for locating loved ones can be misused for exploitable behavior when the requirements for consent \citep{AAAI-22:consent} are violated. Hence, they are dual-use \cite{IPVspyware2018}. 
Through a qualitative study, \citet{Stalkerparadise2018} found that many find-my-phone applications are intended for anti-theft and safety purposes but are heavily misused against privacy. \citet{IPVspyware2018} categorize such apps as spyware, especially in the case of intimate partner surveillance. Just because an app has some legitimate uses doesn't justify its possible misuse. Hence, we consider such apps to be exploitable. Relying on app reviews highlights both types of exploitable functionalities (always malicious and possibly with legitimate uses) against which future and current app users should be warned. Moreover, app distribution platforms and developers should mitigate privacy risks.

\begin{figure}[t]
\centering
\subfigure[The SaferKid \cite{Saferkid} app provides multiple ways to monitor victim's phone activities.]{
\includegraphics[width=0.5\columnwidth]{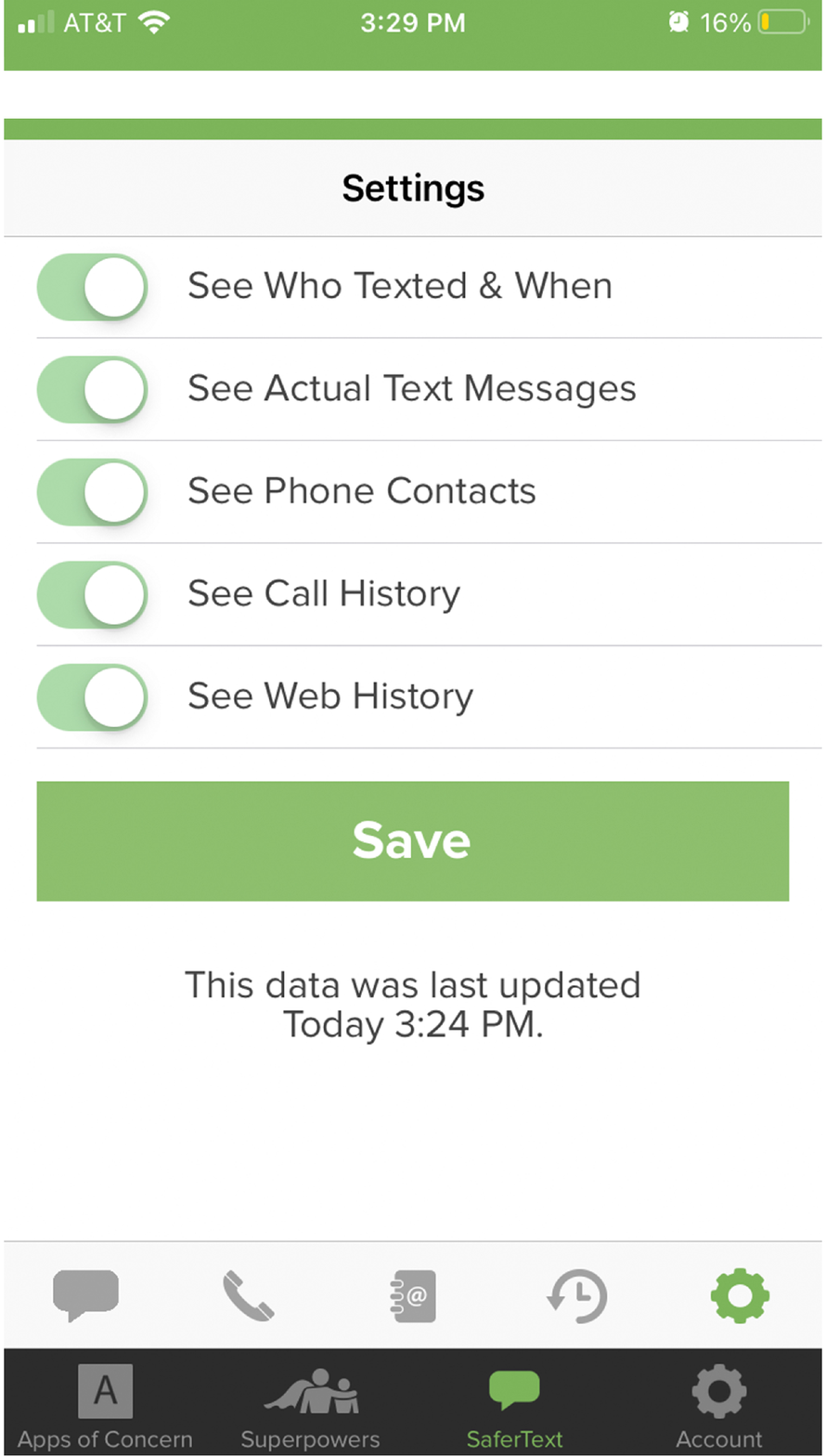}\label{fig:Safercapabilities}
}
\subfigure[Using the SaferKid \cite{Saferkid} app, victim's chats can be seen on the abuser's phone.]{
\includegraphics[width=0.5\columnwidth]{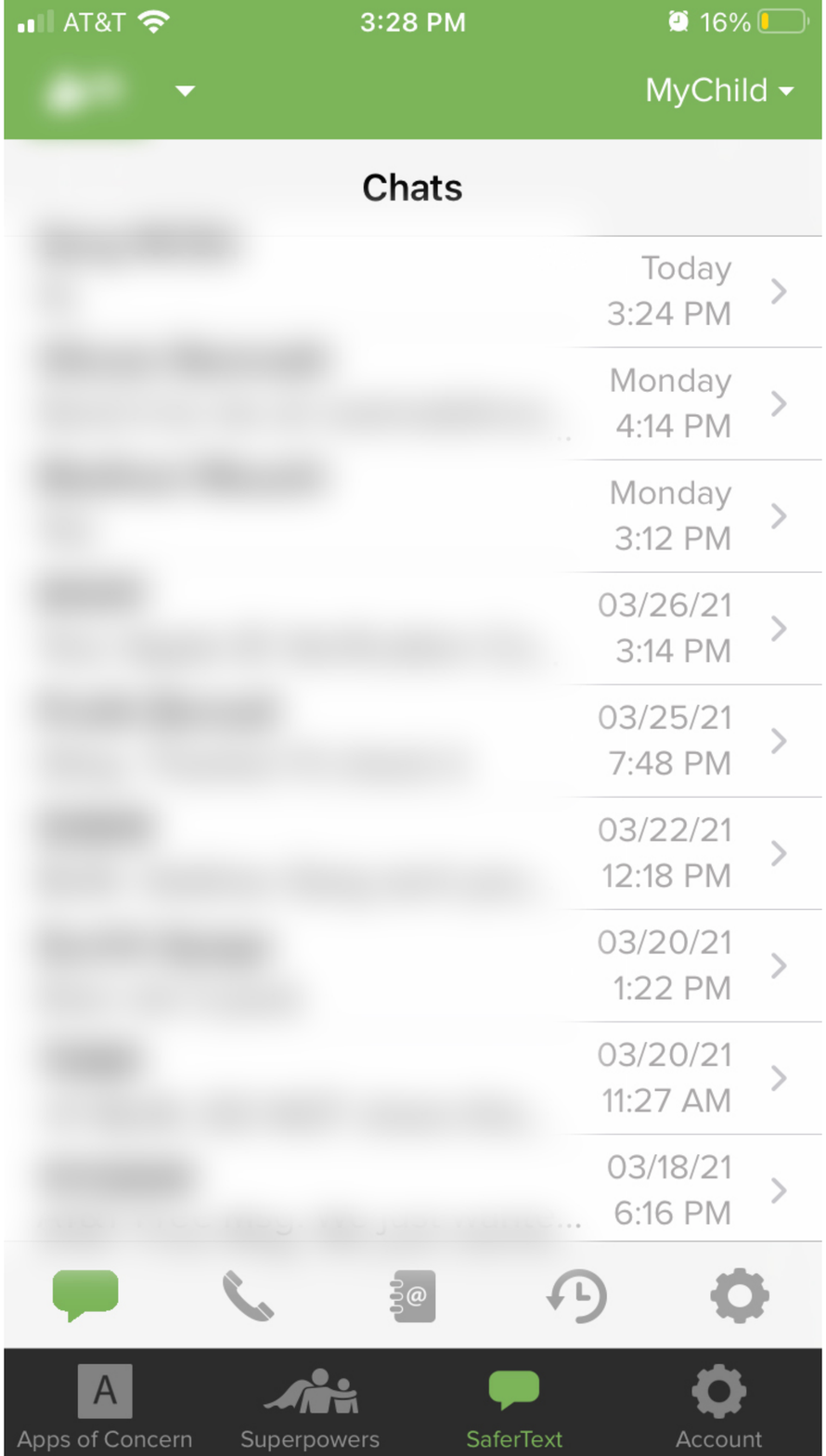}\label{fig:Saferchats}
}

\caption{Exploitable functionalities in SaferKid Text Monitoring \cite{Saferkid}. The actual chats and device's name are hidden for anonymity.}
\label{fig:Appusage}
% \Description{This picture shows two screenshots of the SaferKid Text Monitoring App. The first screenshot shows how this app enables the user (on the ``parent's'' device) to select what activities they wish to monitor on the ``child's'' device. These activities include chat history, chat logs, phone contacts, call history, and Web history. 
% The second screenshot shows how the child's chats can be viewed on the parent's device.}
\end{figure}

We illustrate the exploitable behavior of the SaferKid Text Monitoring App \cite{Saferkid} by installing it on two devices: a parent's device (iOS version 14.4.1) and a child's device (Android version 11.0). 
Figure~\ref{fig:Appusage} shows the exploitable features present in this app. 
Activities on the child's device can be monitored on the synced parent's device. Figure~\ref{fig:Safercapabilities} shows SaferKid exploitable functionalities such as monitoring text messages, web history, and call history. 
We verified each of these functionalities. 
Figure~\ref{fig:Saferchats} shows the screen displaying all chats of the victim. 
Apps such as SaferKid are advertised as safety apps for children but can be secretly or forcefully installed on another device to monitor the user's activity. 
Not only parents but anyone can misuse such apps by installing them on a victim's phone.

\begin{figure}[t]
\centering
\subfigure[Find My iPhone \cite{Findmyiphone} app enables the abuser to track multiple devices at a time.]{
\includegraphics[width=0.5\columnwidth]{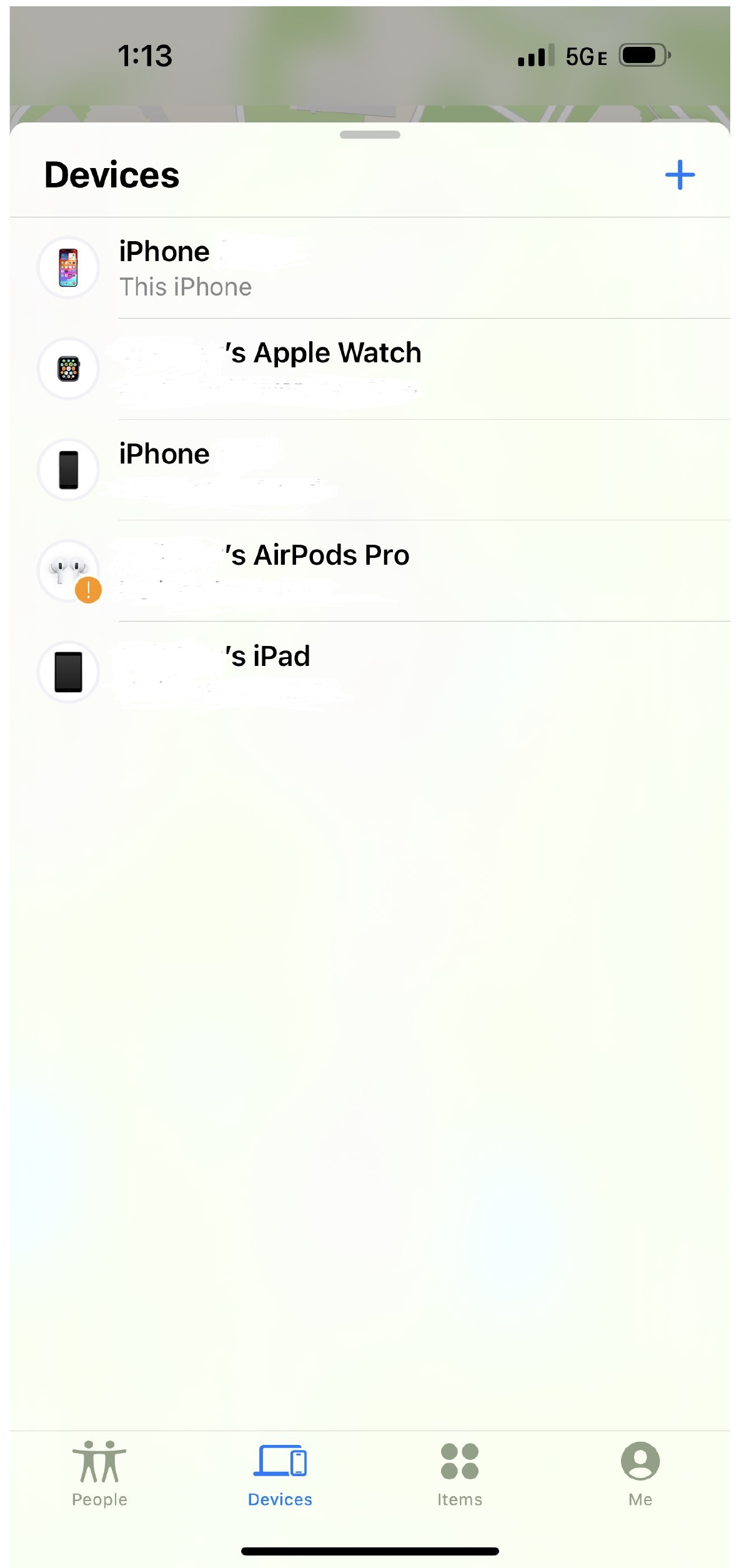}
}
\subfigure[After selecting a device, abuser can see the directions to victim's location.]{
\includegraphics[width=0.5\columnwidth]{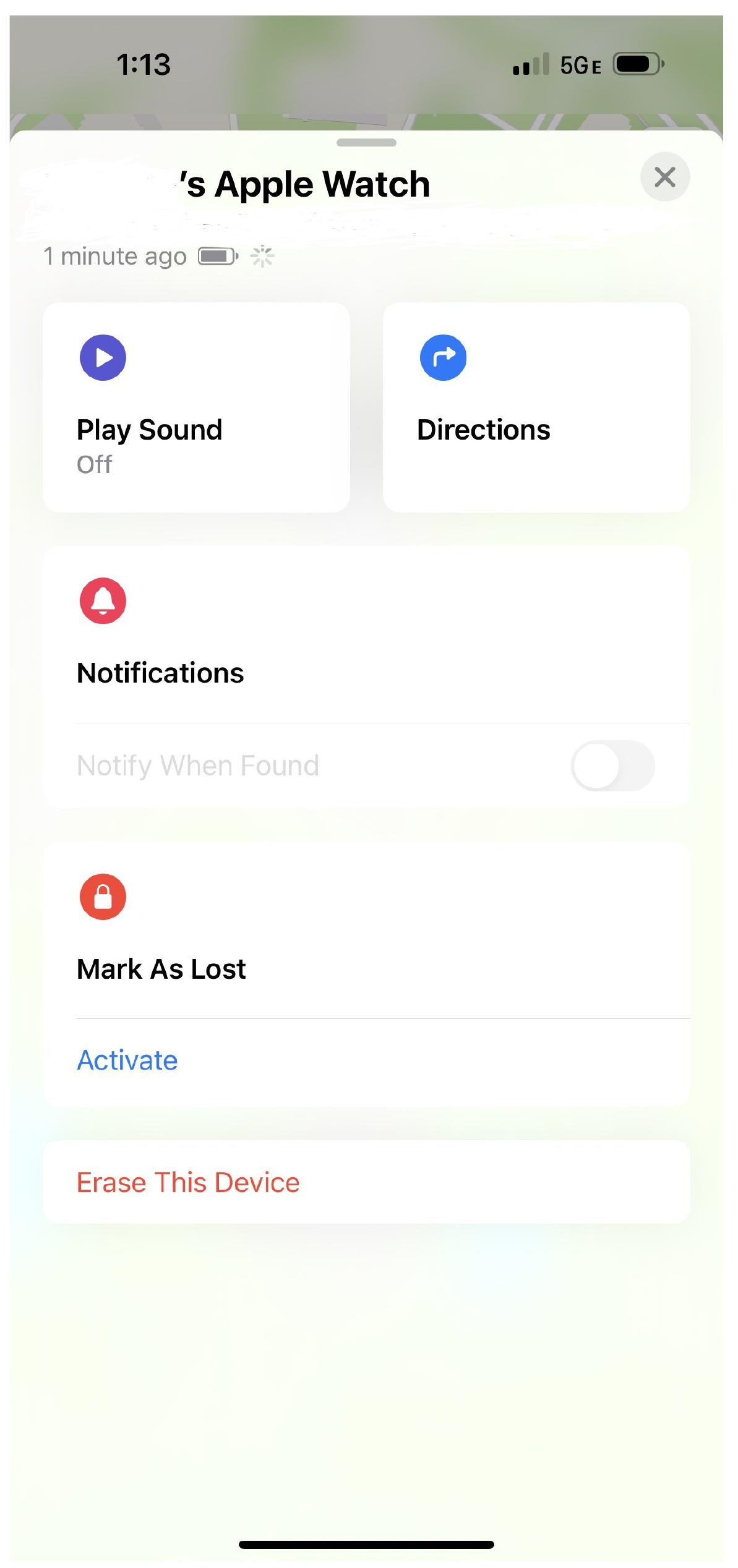}
}

\caption{Exploitable functionalities in Find My iPhone \cite{Findmyiphone}. The names and locations of devices are hidden for anonymity.}
\label{fig:findmyiphone}
% \Description{This picture shows two screenshots of the Find My iPhone App. The first screenshot shows how this app enables the user to track multiple devices (such as an Apple Watch) and the second screenshot shows how the app enables the user to get directions to a connected device.}
\end{figure}

\begin{table*}[!htb]
\centering
\caption{Types of exploitable functionalities.} 
\label{tab:exploitable-capability-types}
% \Description{This table shows examples of the four types of exploitable functionalities we found. The first column lists the exploitable functionality, the second column lists some sample apps that exhibit that functionality, and the third column shows snippets of reviews indicating that functionality.}
\begin{tabu}{X[2.1,L] X[2.1,L] X[8.5,L]}
\toprule
    Exploitable Functionality & App Example & Alarming Review\\
    
    \midrule
    
    Monitoring phone activities& SaferKid Text Monitoring App \cite{Saferkid}  &\fsl{\ldots Tracking things like social media, texts, and search history is just a complete disregard of privacy. You have to have trust in your kids \ldots Apps like these shouldn't be allowed. IF YOU TRUST YOUR KID, DONT DOWNLOAD.} (Date: 2019-12-07)\\

    \midrule

    Audio or video surveillance & Find My Kids: Parental control \cite{Findkids} & \fsl{This app proves to have a invasion of privacy. Due to the fact if your kids was at a friends house and talking to his friends parents, this app records what is going on and is a invasion of privacy. If your child left their phone downstairs or anywhere and they are playing it can record private conversation between adult and is a unsafe \ldots} (Date: 2019-01-16)\\
    
    \midrule
    Tracking location& Find My iPhone \cite{Findmyiphone} &\fsl{\ldots It's supposed to be used to recover a lost phone, not to religiously stalk your children.\ldots The fact that a mom actually installed this app onto her son's phone without his knowledge is flat out wrong. \ldots If you're constantly monitoring your child 24/7, just imagine what your child will do when they go off to college. \ldots} (Date: 2014-02-13)\\

    \midrule
    Profile stalking& Kik Messaging and Chat App \cite{Kik}  &\fsl{MAKE THIS SAFE PEOPLE WANT TO USE IT BUT DON'T WANT TO BE STALKED OR ABUSED IN THE APP PROTECT THE APP OR PEOPLE WON'T USE IT MY FRIEND HAD A MAN SEND HER A BAD PICTURE IF YOU KNOW WHAT I MEAN. OVERALL THIS APP IS GREAT!!!!!"} (Date: 2015-06-24)\\
    
    \bottomrule
    \end{tabu}
\end{table*}

\section{Related Work}
\label{sec:background}
We describe previous works focusing on auditing apps, app reviews, and spying through mobile apps.

\subsection{Auditing and App Reviews}
Most research on software audits analyzes the flow and dependencies in source code \citep{Bluejay2021, EffectiveAudits2019, Vccfinder2015, Vuln4real2020}. Only a few audit studies consider data such as system traces \cite{Mauditor2015}. However, all these studies fail to identify exploitable apps because they focus on the technical aspects of building or running them. Thus, their analysis doesn't address cases of forced consent and inappropriate access to public information. Our work leverages app reviews, unexplored by prior studies, for misuse audits.

Prior studies show that app reviews are valuable in other ways. \citet{Appreviews2018} mine requirement-related information from app reviews and propose an active-learning framework to fetch relevant ones in a semi-supervised manner. \citet{Pixie2022} leverage reviews to draw implicit and explicit comparisons between competing apps. \citet{Casper2020} extract $\langle$action, app problem$\rangle$ pairs from the stories expressed in reviews, which can help direct developers to common problems. \citet{ARMiner2014} showcase the most relevant reviews to developers by grouping informative ones and applying unsupervised techniques. For developers to learn from feature suggestions and bug reports, \citet{UserRationale2017} filter rationale-backed reviews through classification. Some studies \citep{Userreviews2019, Userperception2020} mine reviews to understand users' perceptions about apps selling their data. However, app reviews remain largely unexplored for understanding misuse by an app's users.

\subsection{Spying through Mobile Apps}
Chatterjee \etal \cite{IPVspyware2018} apply search queries and manual verification based on information such as app descriptions and permissions to identify Intimate Partner Surveillance apps. 
However, in general, the actual usage deviates from the legitimate purpose shown in app descriptions. 
To identify such misuse, we focus on the evidence provided in app reviews.

Roundy \etal \cite{Creepware2020} identify apps used for phone number spoofing and message bombing, which lie outside our scope of exploitable apps. Conversely, exploitable apps include those that enable stalking public information, which lie outside their scope. 
Roundy \etal use installation data, to uncover spying apps that are installed on infected devices. However, we focus on evidence of exploitable behavior present in app reviews to uncover exploitable apps. Roundy \etal rely upon Norton's security app \cite{NortonApp} to determine which devices are infected. Thus, their approach would miss apps that a general user can leverage to spy.

All these studies, along with a few others \citep{IPSTools2020, IPVcutomersupport2021,Safetydilemma2021}, investigate how technology is abused for spying. However, they do not support the broader set of exploitable apps. In particular, they do not consider cases when the victim is uncomfortable with the access (e.g., of public information) even if aware of it, along with cases involving family, not only strangers, as abusers. 

Some studies address user privacy in the context of user information shared with software developers \cite{Leakdata2019,Covertchannel2017,Colludingapplications2012} . 
They do not consider cases where an app user can exploit the app to access the information of a victim (another app user or bystander).

\section{Discussion}
\label{sec:conclusion}
%%%%%%%%%%%%%%%%%%%%%%%%%%%%%%%%%%%%%%%%%%%%%%%%%

We proposed \approach, an approach to automatically analyze app reviews for misuse audits. Specifically, \approach identifies exploitable apps along with their exploitable functionalities from reviews. Doing so reduces the burden of manually installing all apps to perform a misuse audit. \approach predicts exploitable apps from multiple sources. After verification of reviews and apps' descriptions, we confirmed \np{156} apps facilitating the misuse. To confirm their misuse, we manually investigated their reviews in two steps. First, we scrutinized their top \np{50} alarming reviews to confirm any exploitable functionalities. Second, we scrutinized their reviews containing our keywords. Solid misuse cases (in at least one of these two steps) give credence to the exploitability of these \np{156} apps (full list in the appendix ordered alphabetically).

Below, we describe our process of reporting exploitable apps, threats to validity, and promising future directions.

\subsection{Reporting to Apple App Store and App Developers}

We informed the Apple App Store about the potential misuse of all exploitable apps that we confirmed above. We found that \np{17} of them have already been deleted from the Apple Store, possibly because of raised privacy concerns or other unknown reasons. 

Only \np{90} developers (of apps that we confirmed as exploitable) provided their email or chat support on the Apple App Store. We contacted these developers and informed them about the misuse cases (template in appendix). We heard back from \np{14} of them. Despite asking specific questions, we received generic replies from them. Six of them acknowledged the misuse or informed that their privacy teams take necessary steps in whatever way possible; Five developers assumed that app users (including abusers) would never misuse their apps (an unrealistic assumption); responses from two developers suggested they were not receptive to our feedback; remaining one developer did not answer back after we asked more specific questions.

\subsection{Threats to Validity}

We now discuss the threats we identify in our work. The identified threats are of two types: (i) the threats that we mitigate; and (ii) the threats that remain.

\subsubsection{Threats Mitigated}
\label{sec:threats-mitigated}
We mitigated the following threats to validity. To address a potential bias because of our keywords, we constructed a training set that was balanced between reviews that matched and didn't match our keywords. Using this training set helped our model learn from the context and not from specific keywords. Second, instead of crowd workers, who might not understand the problem well, two authors of this paper annotated the whole training data. Third, the ground truth (of exploitable apps) could be biased if it was formed only using top alarming reviews. We mitigated this bias by scrutinizing reviews from seed and snowball datasets (explained in appendix) containing our keywords, which can have evidence missed by alarming reviews.

\subsubsection{Threats Remaining}
Now, we describe the threats that remain in our work. 
First, we investigated only a few thousand apps, which may not represent all apps on the Apple App Store \cite{Appstore}. The performance of \approach may be reduced while testing it on all apps of the Apple App Store. 
Second, we targeted apps and their reviews only on Apple's App Store. Upon deployment on other app stores, the performance of our approach can differ. Third, suppose an app distribution platform does not offer recommendations for similar apps. In that case, finding good candidates for exploitable apps will be a challenge. An alternative is to prioritize applying \approach for the apps flagged by app users (victims). Fourth, some reviews in our datasets are old. Due to frequent app updates, the types of misuse may change over time. Although we confirm that our findings are still relevant in Section~\ref{sec:relevance}, there is a possibility that a few apps may not be exploitable anymore. Fifth, some findings are based on a random review sample because manually reading all reviews is not feasible. The results may vary in the latter case.

\subsection{Limitations and Future Directions}

We identify the following limitations of this approach, each of which suggests topics for future research. 
First, \approach may miss some exploitable apps if they do not have alarming reviews at the time of analysis, possibly because they are new apps. 
However, this limitation can be overcome if \approach is used along with app descriptions. Users could be warned against apps whose descriptions are similar to already identified exploitable apps (or against similar apps found through a recommendation system). Further, 
\approach can flag these new apps as soon as their alarming reviews arrive. In this way, our computational model can be updated by including newly found exploitable cases in the training set and iteratively identifying a large landscape of misuse. 

Second, uncovering exploitable functionalities involves the manual effort of inspecting top alarming reviews. A future direction is to automate this process. Moreover, each functionality is analyzed in isolation from other functionalities. Future audit studies can work on the interdependence of the functionalities facilitating misuse. 

Third, app reviews describe perceptions of an app and its misuse, which can be affected by their personal preferences, e.g., how much information access is too much to share. Moreover, cases involving family power dynamics (parent-child and intimate partners), especially regarding forced consent, are nontrivial to analyze user interactions and perceptions. Manual inspection of such reviews by developers can mitigate but not eliminate this limitation. Future studies can explore recommending apps based on a user's personal preferences. 

Fourth, some reviews may falsely claim the exploitable behavior of the app. This includes reviews written by the app's competitors or by people assuming themselves victims. Identifying such misleading cases is challenging and out of the scope of our study. 

Fifth, app reviews can be ambiguous in differentiating between inherently malicious and benign but exploitable functionalities. For example, the Safer Kid Text Monitoring App \cite{Saferkid} is designed to keep loved ones safe (benign app) but is misused according to reviews. Future research can distinguish between these benign apps and other malicious ones.

\section{Reproducibility}
\label{sec:reproducibility}

We have released our entire training data along with the computational model on Zenodo \citep{Garg+24:misuse-Zenodo}. Our appendix provides details of (1) dataset collection and annotation, (2) training and evaluation, and (3) linguistic analysis. 

\section*{Acknowledgement}
Thanks to the anonymous reviewers for their helpful comments. 
Thanks to the Department of Defense for partially supporting this research under the Science of Security Lablet.
NA acknowledges partial support from EPSRC (EP/W025361/1).

\bibliographystyle{plainnat}
\bibliography{Hui,Nirav,Vaibhav,Munindar}

\DeclareRobustCommand{\nUmErAL}[1]{#1}\DeclareRobustCommand{\nAmE}[3]{#3}
\begin{thebibliography}{54}
\providecommand{\natexlab}[1]{#1}
\providecommand{\url}[1]{\texttt{#1}}
\expandafter\ifx\csname urlstyle\endcsname\relax
  \providecommand{\doi}[1]{doi: #1}\else
  \providecommand{\doi}{doi: \begingroup \urlstyle{rm}\Url}\fi

\bibitem[Basak et~al.(2007)Basak, Pal, and Patranabis]{Supportvector2007}
Debasish Basak, Srimanta Pal, and Dipak Patranabis.
\newblock Support vector regression.
\newblock \emph{Neural Information Processing – Letters and Reviews}, 11, November 2007.

\bibitem[Besmer et~al.(2020)Besmer, Watson, and Banks]{Userperception2020}
Andrew~R. Besmer, Jason Watson, and M.~Shane Banks.
\newblock Investigating user perceptions of mobile app privacy: An analysis of user-submitted app reviews.
\newblock \emph{International Journal of Information Security and Privacy (IJISP)}, 14\penalty0 (4):\penalty0 74--91, October 2020.

\bibitem[Block et~al.(2017)Block, Narain, and Noubir]{Covertchannel2017}
Kenneth Block, Sashank Narain, and G.~Noubir.
\newblock An {A}utonomic and {P}ermissionless {A}ndroid {C}overt {C}hannel.
\newblock \emph{Proceedings of the 10th ACM Conference on Security and Privacy in Wireless and Mobile Networks}, pages 184--194, 2017.

\bibitem[Borchani et~al.(2015)Borchani, Varando, Bielza, and Larra\~{n}aga]{Mutioutputregression2015}
Hanen Borchani, Gherardo Varando, Concha Bielza, and Pedro Larra\~{n}aga.
\newblock A survey on multi-output regression.
\newblock \emph{Wiley Interdisciplinary Reviews Data Mining and Knowledge Discovery}, 5\penalty0 (5):\penalty0 216--233, September 2015.

\bibitem[Cer et~al.(2018)Cer, Yang, Kong, Hua, Limtiaco, John, Constant, Guajardo{-}Cespedes, Yuan, Tar, Sung, Strope, and Kurzweil]{UniversalSentenceEncoder2018}
Daniel Cer, Yinfei Yang, Sheng{-}yi Kong, Nan Hua, Nicole Limtiaco, Rhomni~St. John, Noah Constant, Mario Guajardo{-}Cespedes, Steve Yuan, Chris Tar, Yun{-}Hsuan Sung, Brian Strope, and Ray Kurzweil.
\newblock Universal sentence encoder.
\newblock \emph{CoRR}, abs/1803.11175:\penalty0 1--7, 2018.

\bibitem[Chatterjee et~al.(2018)Chatterjee, Doerfler, Orgad, Havron, Palmer, Freed, Levy, Dell, McCoy, and Ristenpart]{IPVspyware2018}
Rahul Chatterjee, Periwinkle Doerfler, Hadas Orgad, Sam Havron, Jackeline Palmer, Diana Freed, Karen Levy, Nicola Dell, Damon McCoy, and Thomas Ristenpart.
\newblock The spyware used in intimate partner violence.
\newblock In \emph{Proceedings of the 39th IEEE Symposium on Security and Privacy (SP)}, pages 441--458, San Francisco, CA, USA, May 2018. IEEE Press.

\bibitem[Chen et~al.(2014)Chen, Lin, Hoi, Xiao, and Zhang]{ARMiner2014}
Ning Chen, Jialiu Lin, Steven C.~H. Hoi, Xiaokui Xiao, and Boshen Zhang.
\newblock Ar-miner: Mining informative reviews for developers from mobile app marketplace.
\newblock In \emph{Proceedings of the 36th International Conference on Software Engineering}, pages 767--778, New York, May 2014. Association for Computing Machinery.

\bibitem[Cohen(1988)]{Cohen-88-Statistics}
Jacob Cohen.
\newblock \emph{Statistical Power Analysis for the Behavioral Sciences}.
\newblock Lawrence Erlbaum Associates, Hillsdale, New Jersey, 2nd edition, 1988.

\bibitem[Devine et~al.(2021)Devine, Koh, and Blincoe]{USEMotivation2021}
Peter Devine, Yun~Sing Koh, and Kelly Blincoe.
\newblock Evaluating {U}nsupervised {T}ext {E}mbeddings on {S}oftware {U}ser {F}eedback.
\newblock In \emph{Proceedings of the IEEE 29th International Requirements Engineering Conference Workshops (REW)}, pages 87--95, Indiana, 2021. IEEE.
\newblock \doi{10.1109/REW53955.2021.00020}.

\bibitem[Dhinakaran et~al.(2018)Dhinakaran, Pulle, Ajmeri, and Murukannaiah]{Appreviews2018}
Venkatesh~T. Dhinakaran, Raseshwari Pulle, Nirav Ajmeri, and Pradeep~K. Murukannaiah.
\newblock App {R}eview {A}nalysis via {A}ctive {L}earning: {R}educing {S}upervision {E}ffort without {C}ompromising {C}lassification {A}ccuracy.
\newblock In \emph{Proceedings of 26th IEEE International Requirements Engineering Conference (RE)}, pages 170--181, Banff, Canada, 2018. IEEE.

\bibitem[Flow(2023)]{Tensorflowhub}
Tensor Flow.
\newblock {T}ensorflow {H}ub, 2023.
\newblock URL \url{https://www.tensorflow.org/hub}.
\newblock Accessed 2023-08-05.

\bibitem[Freed et~al.(2018)Freed, Palmer, Minchala, Levy, Ristenpart, and Dell]{Stalkerparadise2018}
Diana Freed, Jackeline Palmer, Diana Minchala, Karen Levy, Thomas Ristenpart, and Nicola Dell.
\newblock ``{A Stalker’s Paradise}'': How intimate partner abusers exploit technology.
\newblock In \emph{Proceedings of the CHI Conference on Human Factors in Computing Systems}, pages 1--13, New York, April 2018. Association for Computing Machinery.

\bibitem[Freed et~al.(2019)Freed, Havron, Tseng, Gallardo, Chatterjee, Ristenpart, and Dell]{Phonehacked2019}
Diana Freed, Sam Havron, Emily Tseng, Andrea Gallardo, Rahul Chatterjee, Thomas Ristenpart, and Nicola Dell.
\newblock Is my phone hacked? analyzing clinical computer security interventions with survivors of intimate partner violence.
\newblock \emph{Proceedings of the 17th ACM Conference on Human-Computer Interaction}, 3:\penalty0 202:1--202:24, 2019.

\bibitem[Garc{\'\i}a et~al.(2021)Garc{\'\i}a, Guerrero, Zeitsoff, Korlakunta, Fernandez, Fox, and Ruiz-Cort{\'e}s]{Bluejay2021}
C{\'e}sar Garc{\'\i}a, Alejandro Guerrero, Joshua Zeitsoff, Srujay Korlakunta, Pablo Fernandez, Armando Fox, and Antonio Ruiz-Cort{\'e}s.
\newblock Bluejay: a cross-tooling audit framework for agile software teams.
\newblock In \emph{Proceedings of 43rd IEEE International Conference on Software Engineering: Software Engineering Education and Training (ICSE-SEET)}, pages 283--288, Online, 2021. IEEE.

\bibitem[Garg et~al.(2024)Garg, Guo, Ajmeri, Bhattacharya, and Singh]{Garg+24:misuse-Zenodo}
Vaibhav Garg, Hui Guo, Nirav Ajmeri, Saikath Bhattacharya, and Munindar~P. Singh.
\newblock Missauditor {D}ataset and {S}oftware, 2024.
\newblock Zenodo {P}ublic {R}epository. Accessed 2024-07-13.

\bibitem[Guo and Singh(2020)]{Casper2020}
Hui Guo and Munindar~P. Singh.
\newblock Caspar: {E}xtracting and {S}ynthesizing {U}ser {S}tories of {P}roblems from {A}pp {R}eviews.
\newblock In \emph{Proceedings of the IEEE 42nd International Conference on Software Engineering}, pages 628–--640, New York, NY, USA, July 2020. Association for Computing Machinery.
\newblock ISBN 9781450371216.
\newblock \doi{10.1145/3377811.3380924}.

\bibitem[Hallgren(2012)]{Hallgren-TQMP12-IRR}
Kevin~A. Hallgren.
\newblock Computing inter-rater reliability for observational data: An overview and tutorial.
\newblock \emph{Tutorials in Quantitative Methods for Psychology}, 8\penalty0 (1):\penalty0 23--34, 2012.

\bibitem[Haque et~al.(2022)Haque, Garg, Guo, and Singh]{Pixie2022}
Amanul Haque, Vaibhav Garg, Hui Guo, and Munindar~P Singh.
\newblock Pixie: {P}reference in {I}mplicit and {E}xplicit {C}omparisons.
\newblock In \emph{Proceedings of the 60th Annual Meeting of the Association for Computational Linguistics}, volume~2, pages 106--112, Dublin, Ireland, 2022. Association for Computational Linguistics.

\bibitem[Havron et~al.(2019)Havron, Freed, Chatterjee, McCoy, Dell, and Ristenpart]{Clinicalsecurity2019}
Sam Havron, Diana Freed, Rahul Chatterjee, Damon McCoy, Nicola Dell, and Thomas Ristenpart.
\newblock Clinical computer security for victims of intimate partner violence.
\newblock In \emph{Proceedings of the 28th USENIX Security Symposium}, pages 105--122, Santa Clara, January 2019. USENIX Association.

\bibitem[{Kurtanovi\'c} and {Maalej}(2017)]{UserRationale2017}
Zijad {Kurtanovi\'c} and Walid {Maalej}.
\newblock Mining user rationale from software reviews.
\newblock In \emph{Proceedings of the 25th IEEE International Requirements Engineering Conference (RE)}, pages 61--70, Lisbon, Portugal, September 2017. IEEE Press.

\bibitem[Manning(2011)]{POS2011}
Christopher~D. Manning.
\newblock Part-of-speech tagging from 97{\%} to 100{\%}: Is it time for some linguistics?
\newblock In \emph{Proceedings of the Computational Linguistics and Intelligent Text Processing}, pages 171--189, Berlin, Heidelberg, February 2011. Springer.

\bibitem[Marforio et~al.(2012)Marforio, Ritzdorf, Francillon, and Capkun]{Colludingapplications2012}
Claudio Marforio, Hubert Ritzdorf, Aur\'{e}lien Francillon, and Srdjan Capkun.
\newblock Analysis of the {C}ommunication between {C}olluding {A}pplications on {M}odern {S}martphones.
\newblock In \emph{Proceedings of the 28th Annual Computer Security Applications Conference}, pages 51--60, New York, December 2012. Association for Computing Machinery.

\bibitem[Mikolov et~al.(2013)Mikolov, Sutskever, Chen, Corrado, and Dean]{Word2Vec}
Tomas Mikolov, Ilya Sutskever, Kai Chen, Greg Corrado, and Jeffrey Dean.
\newblock Distributed representations of words and phrases and their compositionality.
\newblock In \emph{Proceedings of the 26th International Conference on Neural Information Processing Systems - Volume 2}, NIPS, pages 3111--3119, Lake Tahoe, Nevada, December 2013. Neural Information Processing Systems Foundation.

\bibitem[Miller(1995)]{Wordnet1995}
George~A. Miller.
\newblock Wordnet: A lexical database for english.
\newblock \emph{Communications of the ACM}, 38\penalty0 (11):\penalty0 39--41, November 1995.
\newblock \doi{10.1145/219717.219748}.

\bibitem[Mohammad(2018)]{VAD-ACL2018}
Saif~M. Mohammad.
\newblock Obtaining {R}eliable {H}uman {R}atings of {V}alence, {A}rousal, and {D}ominance for 20,000 {E}nglish {W}ords.
\newblock In \emph{Proceedings of The Annual Conference of the Association for Computational Linguistics (ACL)}, Melbourne, Australia, 2018.

\bibitem[{Nguyen} et~al.(2019){Nguyen}, {Derr}, {Backes}, and {Bugiel}]{Userreviews2019}
Duc~Cuong {Nguyen}, Erik {Derr}, Michael {Backes}, and Sven {Bugiel}.
\newblock Short text, large effect: Measuring the impact of user reviews on android app security \& privacy.
\newblock In \emph{Proceedings of the 40th IEEE Symposium on Security and Privacy (SP)}, pages 555--569, San Francisco, May 2019. IEEE Computer Society.

\bibitem[Pashchenko et~al.(2020)Pashchenko, Plate, Ponta, Sabetta, and Massacci]{Vuln4real2020}
Ivan Pashchenko, Henrik Plate, Serena~Elisa Ponta, Antonino Sabetta, and Fabio Massacci.
\newblock Vuln4real: A methodology for counting actually vulnerable dependencies.
\newblock \emph{IEEE Transactions on Software Engineering}, 48\penalty0 (5):\penalty0 1592--1609, 2020.

\bibitem[Pennington et~al.(2014)Pennington, Socher, and Manning]{Glove2014}
Jeffrey Pennington, Richard Socher, and Christopher Manning.
\newblock {G}lo{V}e: Global vectors for word representation.
\newblock In \emph{Proceedings of the Conference on Empirical Methods in Natural Language Processing ({EMNLP})}, pages 1532--1543, Doha, Qatar, October 2014. Association for Computational Linguistics.

\bibitem[Perl et~al.(2015)Perl, Dechand, Smith, Arp, Yamaguchi, Rieck, Fahl, and Acar]{Vccfinder2015}
Henning Perl, Sergej Dechand, Matthew Smith, Daniel Arp, Fabian Yamaguchi, Konrad Rieck, Sascha Fahl, and Yasemin Acar.
\newblock {VCCF}inder: {F}inding {P}otential {V}ulnerabilities in {O}pen-{S}ource {P}rojects to {A}ssist {C}ode {A}udits.
\newblock In \emph{Proceedings of the 22nd ACM SIGSAC Conference on Computer and Communications Security}, pages 426--437, New York, NY, USA, 2015. Association for Computing Machinery.

\bibitem[PyDictionary(2023)]{pydictionary}
PyDictionary.
\newblock {D}ictionary in {P}ython, 2023.
\newblock URL \url{https://pypi.org/project/PyDictionary/}.
\newblock Accessed 2023-08-05.

\bibitem[Pythesaurus(2023)]{Pythesaurus}
Pythesaurus.
\newblock {T}hesaurus in {P}ython, 2023.
\newblock URL \url{https://pypi.org/project/py-thesaurus/}.
\newblock Accessed 2023-08-05.

\bibitem[Reardon et~al.(2019)Reardon, Feal, Wijesekera, On, Vallina-Rodriguez, and Egelman]{Leakdata2019}
Joel Reardon, {\'A}lvaro Feal, Primal Wijesekera, Amit Elazari~Bar On, Narseo Vallina-Rodriguez, and Serge Egelman.
\newblock 50 {W}ays to {L}eak {Y}our {D}ata: {A}n {E}xploration of {A}pps{\textquoteright} {C}ircumvention of the {A}ndroid {P}ermissions {S}ystem.
\newblock In \emph{Proceedings of the 28th {USENIX} Security Symposium}, pages 603--620, Santa Clara, August 2019. {USENIX} Association.

\bibitem[Roundy et~al.(2020)Roundy, Mendelberg, Dell, McCoy, Nissani, Ristenpart, and Tamersoy]{Creepware2020}
Kevin~A. Roundy, Paula~Barmaimon Mendelberg, Nicola Dell, Damon McCoy, Daniel Nissani, Thomas Ristenpart, and Acar Tamersoy.
\newblock The {M}any {K}inds of {C}reepware {U}sed for {I}nterpersonal {A}ttacks.
\newblock In \emph{Proceedings of the 41th IEEE Symposium on Security and Privacy (SP)}, pages 753--770, Los Alamitos, May 2020. IEEE Computer Society.

\bibitem[Singh(2022)]{AAAI-22:consent}
Munindar~P. Singh.
\newblock Consent as a foundation for responsible autonomy.
\newblock \emph{Proceedings of the 36th AAAI Conference on Artificial Intelligence (AAAI)}, 36\penalty0 (11):\penalty0 12301--12306, February 2022.
\newblock \doi{10.1609/aaai.v36i11.21494}.
\newblock Blue Sky Track.

\bibitem[Smith et~al.(2013)Smith, Ganesh, and Liu]{Randomforest2013}
Paul~F. Smith, Siva Ganesh, and Ping Liu.
\newblock A comparison of random forest regression and multiple linear regression for prediction in neuroscience.
\newblock \emph{Journal of Neuroscience Methods}, 220:\penalty0 85--91, October 2013.

\bibitem[Store(2023{\natexlab{a}})]{AirBeam}
Apple~App Store.
\newblock {A}ir{B}eam {V}ideo {S}urveillance {A}pp, 2023{\natexlab{a}}.
\newblock URL \url{https://apps.apple.com/us/app/airbeam-video-surveillance/id428767956}.
\newblock Accessed 2023-10-08.

\bibitem[Store(2023{\natexlab{b}})]{Appstore}
Apple~App Store.
\newblock {A}pple {A}pp {S}tore, 2023{\natexlab{b}}.
\newblock URL \url{https://www.apple.com/app-store/}.
\newblock Accessed 2023-11-11.

\bibitem[Store(2023{\natexlab{c}})]{Findkids}
Apple~App Store.
\newblock {F}ind {M}y {K}ids: {P}arental {C}ontrol, 2023{\natexlab{c}}.
\newblock URL \url{https://apps.apple.com/us/app/id994098803}.
\newblock Accessed 2023-10-08.

\bibitem[Store(2023{\natexlab{d}})]{Findmyiphone}
Apple~App Store.
\newblock {F}ind {M}y i{P}hone {A}pp, 2023{\natexlab{d}}.
\newblock URL \url{https://apps.apple.com/us/app/find-my-iphone/id376101648}.
\newblock Accessed 2023-10-08.

\bibitem[Store(2023{\natexlab{e}})]{Kik}
Apple~App Store.
\newblock {K}ik {A}pp, 2023{\natexlab{e}}.
\newblock URL \url{https://apps.apple.com/us/app/kik/id357218860}.
\newblock Accessed 2023-10-08.

\bibitem[Store(2023{\natexlab{f}})]{Life360}
Apple~App Store.
\newblock {L}ife360 {A}pp, 2023{\natexlab{f}}.
\newblock URL \url{https://apps.apple.com/us/app/life360-find-family-friends/id384830320}.
\newblock Accessed 2023-10-08.

\bibitem[Store(2023{\natexlab{g}})]{NortonApp}
Apple~App Store.
\newblock {N}ortan {M}obile {S}ecurity {A}pp, 2023{\natexlab{g}}.
\newblock URL \url{https://buy-static.norton.com/norton/ps/bb/ushard/4up_mnav05w_us_en_fl_tw_branded_mix-n360.html}.
\newblock Accessed 2023-10-08.

\bibitem[Store(2023{\natexlab{h}})]{OurPact}
Apple~App Store.
\newblock {O}urpact {J}r. {C}hild {A}pp, 2023{\natexlab{h}}.
\newblock URL \url{https://apps.apple.com/us/app/id1127917970}.
\newblock Accessed 2023-10-08.

\bibitem[Store(2023{\natexlab{i}})]{Saferkid}
Apple~App Store.
\newblock {S}aferkid {T}ext {M}onitoring {A}pp, 2023{\natexlab{i}}.
\newblock URL \url{https://apps.apple.com/us/app/saferkid-text-monitoring-app/id1143802529}.
\newblock Accessed 2023-10-08.

\bibitem[Store(2023{\natexlab{j}})]{Threefun}
Apple~App Store.
\newblock {3F}un: {T}hreesome \& {S}wingers {A}pp, 2023{\natexlab{j}}.
\newblock URL \url{https://apps.apple.com/app/id1164067996}.
\newblock Accessed 2023-10-08.

\bibitem[Store(2023{\natexlab{k}})]{Utilitiesappstore}
Apple~App Store.
\newblock {P}opular {U}tilities {A}pps, 2023{\natexlab{k}}.
\newblock URL \url{https://apps.apple.com/us/genre/ios-utilities/id6002}.
\newblock Accessed 2023-11-11.

\bibitem[Store(2023{\natexlab{l}})]{Playstore}
Google~Play Store.
\newblock {G}oogle {P}lay {S}tore, 2023{\natexlab{l}}.
\newblock URL \url{https://play.google.com/store/apps}.
\newblock Accessed 2023-02-15.

\bibitem[Tseng et~al.(2020)Tseng, Bellini, McDonald, Danos, Greenstadt, McCoy, Dell, and Ristenpart]{IPSTools2020}
Emily Tseng, Rosanna Bellini, Nora McDonald, Matan Danos, Rachel Greenstadt, Damon McCoy, Nicola Dell, and Thomas Ristenpart.
\newblock The {T}ools and {T}actics {U}sed in {I}ntimate {P}artner {S}urveillance: {A}n {A}nalysis of {O}nline {I}nfidelity {F}orums.
\newblock In \emph{Proceedings of the 29th {USENIX} Security Symposium}, pages 1893--1909, Virtual, August 2020. {USENIX} Association.

\bibitem[Tseng et~al.(2021)Tseng, Freed, Engel, Ristenpart, and Dell]{Safetydilemma2021}
Emily Tseng, Diana Freed, Kristen Engel, Thomas Ristenpart, and Nicola Dell.
\newblock A digital safety dilemma: Analysis of computer-mediated computer security interventions for intimate partner violence during covid-19.
\newblock In \emph{Proceedings of the 2021 Conference on Human Factors in Computing Systems}, pages 1--17, Virtual, 2021. Association for Computing Machinery.

\bibitem[Uysal and Gunal(2014)]{Textpreprocessing2014}
Alper~Kursat Uysal and Serkan Gunal.
\newblock The impact of preprocessing on text classification.
\newblock \emph{Information Processing \& Management}, 50\penalty0 (1):\penalty0 104--112, 2014.

\bibitem[Wang et~al.(2015)Wang, Jin, and Nahrstedt]{Mauditor2015}
Ting-Yu Wang, Haiming Jin, and Klara Nahrstedt.
\newblock mauditor: Mobile auditing framework for mhealth applications.
\newblock In \emph{Proceedings of the 2015 Workshop on Pervasive Wireless Healthcare}, pages 7--12, New York, NY, USA, 2015. Association for Computing Machinery.

\bibitem[Xia et~al.(2015)Xia, Gong, Lyu, Qi, and Liu]{EffectiveAudits2019}
Mingyuan Xia, Lu~Gong, Yuanhao Lyu, Zhengwei Qi, and Xue Liu.
\newblock Effective real-time android application auditing.
\newblock In \emph{Proceedings of the 36th IEEE Symposium on Security and Privacy}, pages 899--914, San Jose, May 2015. IEEE.

\bibitem[Xu et~al.(2005)Xu, Watanachaturaporn, Varshney, and Arora]{Decisiontree2005}
Min Xu, Pakorn Watanachaturaporn, Pramod~K. Varshney, and Manoj~K. Arora.
\newblock Decision tree regression for soft classification of remote sensing data.
\newblock \emph{Remote Sensing of Environment}, 97\penalty0 (3):\penalty0 322--336, 2005.
\newblock \doi{https://doi.org/10.1016/j.rse.2005.05.008}.

\bibitem[Zou et~al.(2021)Zou, McDonald, Narakornpichit, Dell, Ristenpart, Roundy, Schaub, and Tamersoy]{IPVcutomersupport2021}
Yixin Zou, Allison McDonald, Julia Narakornpichit, Nicola Dell, Thomas Ristenpart, Kevin Roundy, Florian Schaub, and Acar Tamersoy.
\newblock The role of computer security customer support in helping survivors of intimate partner violence.
\newblock In \emph{Proceedings of the 30th {USENIX} Security Symposium}, pages 429--446, Virtual, 2021. {USENIX} Association.

\end{thebibliography}

% \clearpage

\appendix

\subsection{Seed Dataset}
\label{sec:datasetdetails}

Chatterjee \etal \cite{IPVspyware2018} identified \np{2707} iOS apps as potential Intimate Partner Surveillance (IPS) apps, i.e., apps used by a person to spy on their intimate partner. Out of these \np{2707} apps, Chatterjee \etal confirmed \np{414} apps to be IPS, using semi-supervised pruning. 

When we collected our data (from July 2008 to January 2020), \np{1687} (out of \np{2707}) apps existed with at least one review on the Apple App Store. This yielded our \emph{seed dataset} containing \np{11.57} million reviews. Out of {1687} apps in the seed dataset, only \np{210} were confirmed IPS by Chatterjee \etal.

\subsection{Our Keywords}
\label{sec:ourkeywords}

For the formation of a list of keywords, we initialized a set
with the words: \fsl{spy}, \fsl{stalk}, and \fsl{stealth} and queried WordNet \cite{Wordnet1995} for their synonyms. We performed the query operation until we didn't find any new word in the set. The resulting set contained keywords: \fsl{spy}, \fsl{stalk}, \fsl{stealth},  \fsl{descry}, \fsl{chaff}, and \fsl{haunt}. However, \fsl{chaff} and \fsl{haunt} weren't relevant for describing exploitable behavior in reviews. Also, \fsl{descry} was present in only two reviews, both of which were irrelevant to exploitable behavior. To expand the set of keywords, we explored other corpora such as PyDictionary \cite{pydictionary} and Thesaurus \cite{Pythesaurus} but did not find synonyms that are widely used in app reviews. Moreover, keywords used in the previous study \cite{IPVspyware2018}, such as \fsl{track} and \fsl{control} bring many false positive reviews. For example, ``I like \ul{tracking} my distance when I walk with my dog.'' and ``\ldots you can also \ul{control} the audio of your mac through the app \ldots I can control music \ul{tracks} without having to touch the computer.'' are not relevant. Thus, our relevant set of keywords reverts to \fsl{spy}, \fsl{stalk}, and \fsl{stealth}.

\subsection{Linguistic Analysis}
\label{sec:linguistic}
To compute valence, arousal, and dominance scores, we leveraged lexicons provided by \citet{VAD-ACL2018}, which contains scores for about \np{20000} English words. First, from each review, we extracted relevant sentences using our keywords. Second, using parts of speech tagging, we extracted adjectives mentioned in the relevant sentences. Since sentences describe the exploitable behavior, adjectives in them represent the reviewer's sentiment. For example, \fsl{``This \ul{horrible} (adj) app spy on me''}. Third, for each type of reviewer, we computed average scores of adjectives that are present in the lexicon.

\subsection{Annotating Reviews and Model Training}
\label{sec:datatraining}
The annotation was conducted in three steps. First, two authors of this paper rated \np{599} of \np{1884} selected reviews according to the initial set of annotation instructions. The initial instructions included definitions (of convincingness and severity scores) and examples corresponding to each point on the Likert scale.
In this step, for each annotator, we calculated the alarmingness scores of reviews using the convincingness and severity ratings. If the alarmingness scores computed for both the annotators were not at least three (median value on Likert scale), or both are not less than three, annotators resolved such cases via discussions. After discussing, the annotators produced the final set of annotation instructions. In the second step, the annotators followed the final instructions and rated 900 reviews. In this step, we computed an Intraclass Correlation Coefficient (ICC) \cite{Hallgren-TQMP12-IRR} to measure inter-rater agreement. We found ICC(3,k) to be 0.9195 for convincingness and 0.9190 for severity, which accounts for excellent agreement. ICC is suitable for Likert scale ratings. Unlike other measures such as Cohen's \cite{Cohen-88-Statistics} kappa, which is based on (all or nothing) agreement, ICC takes into account the magnitude of agreement (or disagreement) to compute inter-rater reliability. In the third step, the remaining reviews were divided among the two annotators so that only one annotator rates each review.

\subsubsection{Extracting Deep Learning Features from App Review}
\label{sec:sentenceencoder}

We obtained the feature vector of each app review as follows:

\begin{description}[leftmargin=1em]

\item[Combine Sentences:] Remove periods in each app review and combine all its sentences to form single sentence.

\item[Text Preprocessing:] Remove all punctuation marks, stop words \cite{Textpreprocessing2014}, and our keywords, the latter because those keywords may correlate with reviews with higher scores and could create bias in the model.

\item[Sentence Embedding:] We leverage the Universal Sentence Encoder (USE) \cite{UniversalSentenceEncoder2018} to extract embeddings for each app review.
USE uses Deep Averaging Network (DAN) to
provide a 512-dimension embedding for a long text. USE is trained on a large variety of natural language tasks with the aim of capturing the context. In our case, 
USE directly provides sentence level embeddings of an app review, by keeping the context intact. However, alternatives such as GLoVe \cite{Glove2014} and Word2Vec \cite{Word2Vec} lose such context. We leverage the pretrained USE network by using Tensorflow Hub \cite{Tensorflowhub}.
% Hence, we leverage USE instead of word embeddings.

\end{description}

We evaluated the performance of three regression models: support vector regression \cite{Supportvector2007}, random forest \cite{Randomforest2013}, and decision tree \cite{Decisiontree2005}, by ten-fold cross-validation on our dataset. To mitigate bias of our keywords, we remove such keywords in the preprocessing step, so that the regression model learns from the the context of the review and not from specific keywords. As shown in Table~\ref{tab:perfmodel}, the Support Vector Regressor (SVR) yields the smallest MSE of \np{0.625} with \np{0.458} standard deviation, so we choose it for the subsequent phases of our approach.

\begin{table}[!htb]
\caption{Performance of three regression models on ten-fold cross validation.}
\label{tab:perfmodel}
\centering
\begin{tabular}{lrr}
\toprule
Regression Model & Average MSE & Standard Deviation \\
\midrule
Decision Tree  & 1.344 & 0.402\\
Random Forest & 0.712& 0.417\\
Support Vector & \textbf{0.625}&0.458\\
\bottomrule
\end{tabular}
\end{table}

For training, we also explored OpenAI's two advanced embeddings: text-embedding-3-small and text-embedding-3-large. However, while using these embeddings, MSE over 10 folds do not significantly differ (p-value is 0.0582 (>0.05) for small embedding and 0.0533 (>0.05) for large embedding) from MSE values of USE based model. Moreover, training OpenAI's embedding-based model is more expensive than training on USE embeddings. OpenAI's embeddings occupy 3 times (1536-dimensional vector for small embedding) and 6 times (3072-dimensional for large embedding) larger space than USE embedding. Hence, we stick to the USE embedding-based SVR model for predictions.

\subsection{Statistical Methods for Identifying and Ranking exploitable Apps}
\label{sec:exploitable}

\begin{description}[leftmargin=1em]
\item[Weighted Mean of Alarmingness:] In general, for a exploitable app, a small proportion of reviews report exploitable behavior. Thus, we need to catch exploitable apps using their few reviews that have high values on the alarmingness scale. Thus, we assign weights to reviews based on their alarmigness, as follows:

\emph{Defining score buckets:} While annotating reviews, we defined levels of convincing reviews (not convincing to extremely convincing) and severe reviews (not severe to extremely severe) on a Likert scale. We also follow same levels on the alarmingness scale (1: not alarming to 4: extremely alarming). We define a score bucket between every consecutive level of alarmingness (not alarming to slightly, slightly alarming to moderately alarming, moderately alarming to extremely alarming). Table~\ref{tab:scorebuckets1} shows how score buckets are formed using levels of alarmingness.

\emph{Assigning weights to score buckets:}
We have 11.57 million reviews in the seed dataset. Based on the alarmingness computation (discussed in main document), we calculated the probability of a review falling in a score bucket. Since, the reviews reporting exploitable behavior are less, probabilities in buckets~2 and 3 are less than that in bucket~1. We take inverse of these probabilities to get the weights for each score bucket. As a result, we assign higher weights to buckets~2 and 3 than to the bucket~1. Table~\ref{tab:scorebuckets1} also shows the weight assigned to each score bucket. 

\begin{table}[!htb]
\caption{Score buckets for alarmingness.}
\label{tab:scorebuckets1}
\centering
% \begin{tabular}{P{1cm}P{2cm}P{2cm}P{2cm}}
% \begin{tabular}{@{}P{2cm}@{~}P{2cm}@{~}P{0.8cm}@{~}S@{}}
\begin{tabular}{P{1.6cm}P{2.6cm}P{0.8cm} n{1}{3}}
\toprule
Alarmingness Score Range & Alarmingness Level Range & Bucket & {Bucket Weight}  \\
\midrule

$[1,2)$& Not alarming to Slightly & 1 & 2.29e-3 \\
$[2,3)$ & Slightly to Moderately & 2 & 6.08e-2\\
$[3,4]$ & Moderately to Extremely & 3 & 9.36e-1\\
\bottomrule
\end{tabular}
\end{table}

If $a\fsub{1}$, $a\fsub{2}$, \ldots, $a\fsub{n}$ are the alarmingness scores of an app's reviews, and $w\fsub{1}$, $w\fsub{2}$, \ldots, $w\fsub{n}$ are their respective weights (according to Table~\ref{tab:scorebuckets1}), then, $W\fsub{alarmingness}$, the weighted mean of alarmingness is given by:
\begin{equation*}
W\fsub{alarmingness}=\frac{a\fsub{1}*w\fsub{1}+ a\fsub{2}*w\fsub{2}+ \ldots a\fsub{n}*w\fsub{n}}{w\fsub{1}+ w\fsub{2}+ \ldots w\fsub{n}}    
\end{equation*}

The weighted mean of alarmingness ranges from 1 to 4.

\item[Normalized Count:] The weighted mean of alarmingness does not account for the count of reviews that report exploitable behavior against an app. Suppose, \emph{app A} has 15 reviews reporting exploitable behavior and \emph{app B} has 25 reviews reporting exploitable behavior. If all reviews reporting exploitable behavior have the same alarmingness score, the weighted means of the two apps would be the same. But, \emph{app B} shows more evidence of exploitable behavior and should have a higher exploitable score than \emph{app A}. Thus, we also consider the count of reviews. For each app, we calculate the number of reviews in bucket~3. We tried incorporating counts of other buckets, but it led to worse performance of the approach. 

The minimum possible value of the count is zero. However, in some cases, counts can be high, leading to no definite upper limit. Thus, we normalize the counts of all the apps between one and four.
\end{description}

We want to assign high exploitable score to apps that have high scores in both (i) weighted mean of alarmingness and (ii) normalized count. Thus, exploitable score is computed as the geometric mean of these two values.

\subsubsection{Selecting Threshold for Prediction}
\label{sec:verification}

For each app in the seed dataset, \approach computes the exploitable score. All apps are ranked in decreasing order of exploitable scores. The apps with a score greater than a threshold are predicted exploitable. To decide the correct threshold, we follow two steps: (i) label the ground truth of exploitable apps and (ii) vary a threshold between certain values and choose the threshold which gives the best performance of \approach.

We create our ground truth by manually scrutinizing reviews. However, scrutinizing reviews of all \np{1687} apps (seed dataset) is not feasible. Thus, first, we scrutinize the most \np{50} alarming reviews (with a minimum score of two---at least slightly alarming) for apps with the highest \np{100} exploitable scores. Second, we scrutinize reviews containing our keywords for these \np{100} apps. The first step is aligned with our approach since it checks the top alarming reviews. However, the second step is neutral because it searches for evidence for the apps that \approach failed to identify through top alarming reviews. This way we mitigate the threat of bias while curating ground truth for apps. On average, top \np{50} alarming reviews cover $\sim$\np{75}\% distribution of these alarmingness scores, hence we decided to scrutinize the most \np{50} alarming reviews for each of these apps. We also adopt this step of review scrutiny for uncovering exploitable functionalities (Section~4 of the main manuscript) and investigating the relevance of our findings (Section~3.2.3 of the main manuscript). 

We label an app as exploitable provided any of the scrutinized reviews report a exploitable behavior. Table~\ref{tab:instruction} shows the types of reviews we consider indicative or otherwise of a exploitable evidence. If the reviews of an app describe information access performed without the victim's knowledge or when the victim shows discomfort (first three reviews in Table~\ref{tab:instruction}), we consider the app as exploitable. Also, some exploitable apps are used for positive purposes such tracking family members for safety (fourth review in Table~\ref{tab:instruction}) but still possess a potential for future misuse. Similarly, apps used for tracking pets or other objects are considered exploitable (fifth review). Reviews of some apps don't possess any misuse in present or in future, leading to their final label of not exploitable (sixth review). Through manual inspection, we determine that of the \np{100} apps, \np{73} are exploitable and \np{27} are not.

\begin{table*}[!htb]
    \centering
    \caption{Instructions followed for labeling exploitable apps.}
    \label{tab:instruction}    
    \begin{tabular}{p{3cm}p{3.3cm}p{9.0cm}c}
    \toprule
        Type of case & Subtype of case & Example & Exploitable? \\
        \midrule
          Tracking people's information& Without the victim's knowledge & \emph{``Now that I can spy on my wife I will always know when she is cheating''}  & Yes
          \\
        
         Tracking people's information & With the victim's knowledge but with discomfort& \emph{``Ok my mom got this for me and \ldots it's kinda creepy that this app was made so parent could basically stalk their kids.''} & Yes \\
         Tracking people's information & Public information but the victim is uncomfortable &   \emph{``I had someone cyberstalking and harassing me. Multiple attempts in every way shape and form were made to contact app-name to block and ban the stalker's account due to a concern for my well- being.''} & Yes \\

          Tracking people's information& Positive purpose & \emph{``I love finding my family members. Wife was in bad car wreak and I was able to find her location using this app. Thank you!''} & Yes\\
        Tracking pets or other objects& & \emph{``Wow! Day one and I'm stalking my puppet like a soccer mom that ran out of adderall! I'm very excited to use this to interact with my puppet while I'm at work and to check in on the dog walker!''} & Yes \\

         Not related to information accessing & & \emph{``an absolutely amazing and very helpful app.  i don't know how i would keep track of prayer times without it.  love the app.  thank u!!!''} & No \\
    \bottomrule
    \end{tabular}
\end{table*}

For the 100 apps, the exploitable score varies between 1.74 and 3.60. We vary the threshold from 1.73 to 3.59 in steps of 0.01. Apps with a exploitable score above the threshold are predicted exploitable but not exploitable otherwise. At each value of threshold, we report the recall, precision, and F1 score.

Table~\ref{tab:thresholding} shows the performance achieved at specific thresholds. As we increase the threshold, the precision increases at the cost of recall. For exploitable apps, a false negative costs more than a false positive because a false negative leaves a exploitable app undetected, which can harm many victims, whereas a false positive causes only wasted effort in manual scrutiny. Hence, achieving high recall is more important than achieving high precision. Thus, from Table~\ref{tab:thresholding}, we choose \np{1.73} threshold that gives the best recall of \np{100}\% at \np{73}\% precision. Since we fine-tune the threshold on the same seed dataset, we also check \approach's performance (using the chosen threshold) on the other dataset (Section~\ref{sec:snowball}).  

\begin{table}[!htb]
\caption{Choosing an appropriate threshold according to the recall scores.}
\label{tab:thresholding}
\centering
% \begin{tabular}{P{1cm}P{2cm}P{2cm}P{2cm}}
% \begin{tabular}{@{}P{2cm}@{~}P{2cm}@{~}P{0.8cm}@{~}S@{}}
\begin{tabular}{rrrr}
\toprule
Threshold & Precision (\%) & Recall (\%)& F1 Score (\%)\\
\midrule
\textbf{1.73}&	\textbf{73.00}&	\textbf{100.00}&	\textbf{84.39}\\
1.74&	73.40&	94.52&	82.63\\
1.75&	76.13&	91.78&	83.22\\
1.76&	76.82&	86.30&	81.29\\
1.77&	76.92&	82.19&	79.47\\
1.78&	78.37&	79.45&	78.91\\
1.79&	79.71&	75.34&	77.46\\
1.80&	80.95&	69.86&	75.00\\
1.81&	80.95&	69.86&	75.00\\
1.82&	81.96&	68.49&	74.62\\
1.83&	81.03&	64.38&	71.75\\
1.84&	80.35&	61.64&	69.76\\
1.85&	82.69&	58.90&	68.80\\
1.86&	83.33&	54.79&	66.11\\
1.87&	82.97&	53.42&	65.00\\
\bottomrule
\end{tabular}
\end{table}

\begin{mybox}[box:seed-exploitable-behavior1]{Exploitable app from seed dataset}
% \begin{center}\textbf{\ul{exploitable App from Seed Dataset}}\end{center}

\textbf{App: Find My Family \& Friends \cite{Life360}} \\ 
\textbf{Exploitable Score: 3.60/4.00}\\

\textbf{Alarming Review 1} \\
\textbf{Alarmingness: 4.00/4.00}\\
Date of Review: 2019-11-28\\
\fsl{``\ldots Such a terrible thing for unaware parents to use. \ul{Most parents think teens don't need privacy and they constantly need to know where they are and what they're doing and who they're with at all times.} This may make the parent feel at peace but what about the child? It's selfish of parents to not take into consideration of how the teen may feel about always having this app and the parent giving them a very stalkish feeling, it's very uncomfortable.''}\bigskip

\end{mybox}

\begin{mybox}[box:seed-exploitable-behavior2]{Exploitable app from seed dataset}
% \begin{center}\textbf{\ul{Apps with No exploitable Behavior}}\end{center}

\textbf{App: OurPact Jr. Child App \cite{OurPact}} \\
\textbf{Exploitable Score: 2.47/4}\\

\textbf{Alarming Review 1} \\
\textbf{Alarmingness: 4.00/4.00}\\
Date of Review: 2018-06-19\\
\fsl{``\ldots \ul{however this app shuts down almost everything and can see every text and website you've visited}. now, i haven't done anything bad online (recently), but \ul{i find that a little creepy and honestly an invasion of privacy. no wonder this app has such crappy reviews}. also, i used to have way more apps than i do now. because my parents now have the ability to restrict apps that may be ``inappropriate''. i already have to ask permission to download apps, so if they were inappropriate my parents wouldn't let me download them. there's too many apps like this and i think kids need a break from all this crap on their devices.''}

\bigskip

\textbf{Alarming Review 2} \\
\textbf{Alarmingness: 4.00/4.00} \\
Date of Review: 2018-08-16\\
\fsl{``This is a useless app that no parent need to install I pray for every child who has this app installed on their electronics some parents don't understand the modern society but that's okay (but not really) I'm only given 2 hours and writing this review is using up time WHICH IS NOT FRIKEN OK!!! \ul{I hate this hate this app and I hope every child that has had their device attacked by this installment hates this app as much as me.} This app should never be okay to use its inappropriate and everybody's children who have this app installed are making there children ANTI-SOCIAL AND VERY NOT COOL. I have many reasons why this app is SOOOOOOO scaring and dreadful so if your reading and thinking about installing this on ur child's device DONT INSTALL IT because that will ruin their future.''}

\end{mybox}

\subsubsection{Performance of Baseline Methods}

On the seed dataset, we also check the performance of baseline methods described below.
\begin{description}[leftmargin=1em]
\item[Our Keywords on App Description.] We search for the presence of one of our keywords (\fsl{spy}, \fsl{stalk}, and \fsl{stealth}) in app descriptions. Apps whose descriptions contain any of these keywords are predicted exploitable, whereas other apps are predicted as not exploitable.

\item[Extended Keywords on App Description.] We identify additional relevant keywords by extracting verbs through Part-Of-Speech (POS) tagging \cite{POS2011}. POS tagging marks every word in a sentence to an appropriate part of speech (verb, noun, adjective, and so on). Applying this process on descriptions of 73 exploitable apps (from the ground truth) produced 145 verbs, out of which six (\fsl{track}, \fsl{monitor}, \fsl{locate}, \fsl{control}, \fsl{stolen}, \fsl{lost}) we selected as relevant to exploitable behavior. The verbs: \fsl{stolen} and \fsl{lost} are relevant because they describe the apps that are used to find a misplaced phone, which indicates an ability to track another device. We extend our keywords by adding these six verbs. Apps whose description contain these keywords are predicted exploitable.

\item[T\% Keyword Reviews.] For each app, we compute the percentage of reviews containing our keywords. We set a threshold, \fsl{T}, on this percentage, above which apps are predicted exploitable. In our evaluation, \fsl{T} takes the values of 0.3, 0.2, and 0.1, respectively.  

\end{description}

Table~\ref{tab:performancecomparison} summarizes the precision, recall, and F1 scores of all baselines and our approach. Our keywords on the description predict only one exploitable app, leading to 100\% precision (highest among all). However, our keywords miss 72 exploitable apps, which leads to the worst recall of 1.36\%. Among all the baselines, keyword search on reviews with 0.1\% threshold achieves the highest recall of \np{65.07}\%, which is much lower than \approach's recall value. \approach's better performance may be due to fine tuning \approach's threshold on the same seed dataset. Thus, we also compare \approach's performance with these baselines on the other dataset (Section~\ref{sec:snowball}).

\begin{table}[!htb]
\caption{Performance (in \%) of baseline methods and \approach on the seed dataset. Bold value for a metric indicates the highest score among all approaches.}
\label{tab:performancecomparison}
\centering
\begin{tabular}{p{4.5cm}ccc}
\toprule
Method & Recall & Precision & F1\\
\midrule
Our keywords on app descriptions & 01.36& \textbf{100.00} & 02.68 \\
Extended keywords on app descriptions & 61.64& 80.35& 69.76 \\
% Reading App Description& 73.97 & 87.09 & 80.00 \\
\midrule
0.3\% keyword reviews & 46.03& 96.66& 62.36 \\
0.2\% keyword reviews& 50.79& 96.96 & 66.66 \\
0.1\% keyword reviews &65.07 & 95.34& 77.35 \\

\midrule
\approach & \textbf{100.00} & 73.00& \textbf{84.39} \\
\bottomrule
\end{tabular}
\end{table}

Examples~\ref{box:seed-exploitable-behavior1} and~\ref{box:seed-exploitable-behavior2} show alarming reviews of Find My Family \& Friends App \cite{Life360} and OurPact Jr. Child App \cite{OurPact}, which \approach correctly identifies as exploitable. Both of them are dual-use apps. Find My Family \& Friends is a safety app, but alarming reviews report parents misusing the tracking functionality on children, to which children are uncomfortable. Moreover, the alarming reviews of the OurPact Jr. Child App report that parents can monitor children's texts and visited websites by installing the app on the child's device. 

\subsection{Performance on Snowball Dataset}
\label{sec:snowball}
For a fair evaluation of \approach on the snowball dataset, we manually scrutinized significantly more apps than what \approach predicted. Hence, using the same labeling process as described in Section~\ref{sec:verification}, we curated the ground truth, for the apps with the highest \np{140} exploitable scores. Out of these \np{140} apps, we label \np{81} apps as exploitable.

Table~\ref{tab:performancecomparisonsnowball} shows the performance of all baseline methods and \approach on the snowball dataset. Our keywords when used on descriptions predict no app as exploitable, leading to the lowest recall. This is because, on Apple App Store \cite{Appstore}, dual-use apps are not advertised using keywords: \fsl{spy}, \fsl{stalk}, and \fsl{stealth}.

On app descriptions, extended keywords perform better (61.72\% recall at 79.36\% precision) than our keywords due to commonly used words (such as \fsl{track}, \fsl{locate}) in app descriptions. Our keywords when applied on reviews yields 100\% precision, however at a low recall value of 30.86\%. Having high recall is desirable in the context of exploitable apps. Our model yields the best recall of 71.6\% as compared to all other baselines.

\begin{table}[!htb]
\caption{Performance (in \%) of baseline methods and \approach on the snowball dataset. Bold value for a metric indicates the highest score among all approaches.}
\label{tab:performancecomparisonsnowball}
\centering
\begin{tabular}{l ccc}
\toprule
Method & Recall & Precision & F1 \\
\midrule
Our keywords on app descriptions & 0& -- & -- \\
Extended keywords on descriptions & 61.72& 79.36& \textbf{69.44} \\
% Reading App Description& 73.97 & 87.09 & 80.00 \\
\midrule
All keyword reviews& 30.86& \textbf{100} & 47.16 \\

\midrule
\approach & \textbf{71.60} & 64.44& 67.84 \\
\bottomrule
\end{tabular}
\end{table}

\subsection{Scrutinizing Utility Apps}
\label{sec:utilities}
We consider \np{100} popular utility apps that are mentioned on the Apple App Store page \cite{Utilitiesappstore}. Out of \np{100} apps, nine are already scrutinized either in the seed or snowball dataset. For the rest \np{91} apps, we retrieved \np{392928} reviews, over the duration of October 2008 to August 2022. 

\approach predicts only one app as exploitable, which after reviews' scrutiny by us, comes out to be non-exploitable. We also scrutinize \np{10} apps with the highest exploitable scores, by reading their top 50 alarming reviews and reviews containing our keywords. But, none of them are actually exploitable.

\subsection{Contacting App Developers}
We used the following template to reach out to app developers.

\begin{mybox}[box:email]{Reaching out to developers.}
We are security and privacy researchers from the NC State University.  We found that your app is designed for legitimate usage but may still be misused against the privacy of some users. We got to know such potential cases through your app reviews shown below. \\

\textless Sample Alarming Reviews\textgreater\\
% \nsa{how many sample reviews?}\vg{Typically 1-3}

We thought this issue is worth bringing to your attention so that you can take appropriate measures. Please let us know what you think about the same by answering the following questions:

1. Were you aware that users may misuse your apps for the wrong purposes?

2. Do you consider redesigning the app to prevent the users from misuse?

3. Any other steps you think can be taken to ensure victims’ privacy?
\end{mybox}

\subsection{List of exploitable Apps Identified}
\label{sec:listofapps}

\begin{inparaitem}
\item 5-0 Radio Police Scanner
\item Ahgoo baby monitor - audio and video monitoring
\item Alfred Home Security Camera
\item AngelSense Guardian
\item Annie Baby Monitor: Nanny Cam
\item AT\&T FamilyMap®
\item AT\&T Secure Family Companion
\item Baby Monitor 3G
\item Baby Monitor 3G/4G/5G/Wi-Fi 
\item Baby Monitor: Video Nanny Cam
\item Baby Tracker by Sprout
\item Bark - Parental Controls
\item BeenVerified: People Search
\item Bloomz: For Teachers \& Schools
\item Bond - Personal Security
\item Boomerang Parental Control
\item bSafe - Never Walk Alone
\item bthere
\item BuddyTag
\item Canopy - Parental Control App
\item Carpin - Find Family \& Friends
\item Circle 1st Generation
\item Circle Parental Controls App
\item Citizen: Local Safety Alerts
\item Cloud Baby Monitor
\item Covenant Eyes: Quit Porn Now
\item Covert Alert
\item Ear Spy: Super Hearing
\item FamiGo: Parental Control App
\item Familo: Find My Phone Locator
\item Family GPS Locator GeoZilla
\item Family GPS Tracker Kid Control
\item Family Locator and GPS Tracker
\item Family Locator by Fameelee
\item Family Locator by Fameelee
\item Family locator My Family
\item Family Orbit: Parental Control
\item FamilyTime - Parental Controls
\item FamilyTime App For Kid's iPhone \& iPad
\item FamilyWall: Family Organizer
\item FamiSafe Jr - Blocksite
\item FamiSafe-Parental Control App
\item Find My Family Friends \& Phone
\item Find My Family, Friends, Phone
\item Find My Friends - Phone Locator
\item Find my Friends \& Family Track
\item Find my Friends, Family Phone
\item Find My iPhone
\item Find My Kids ~ Footprints
\item Find my kids: child GPS tracker
\item Find my Phone - Family Locator
\item Find my Phone, Friends - iMapp
\item Find My Phone, Friends Tracker
\item Findup: Phone Location Tracker
\item FollowMee GPS Location Tracker
\item Funa - Family Locator
\item GeoLoc - GPS Location Tracker
\item GizmoHub
\item Glympse-Share your location
\item GoGuardian Parent App
\item Google Family Link
\item GPS Phone Tracker for Smartphones
\item GPS TRACKER (Phone location tracking)
\item GPS Tracker | GPS tracking
\item GPS Tracker and Locator Chirp
\item GPS Tracker by FollowMee
\item GPSme Friends \& Family Tracker
\item GroupMe
\item Here Comes the Bus
\item HeyTell
\item HiddenApp, Find My Device App
\item Hum: GPS Family Locator
\item iHeartCam Home Security Camera
\item iMap - Find My Phone \& Friends
\item iSharing: Share Live Location
\item Kaspersky Safe Kids with GPS 
\item Kidgy - Parental control app
\item Kidgy: Find My Family
\item Kidslox - Parental Control App
\item Kik Messaging \& Chat App
\item LMK: Make New Friends
\item Locate \& Track Phone By Number
\item Location Tracker - find GPS
\item Locator for Family \& Friends
\item MamaBear Family Safety
\item Meet24 - Flirt, Chat, Singles
\item MeetMe - Meet, Chat \& Go Live
\item Microsoft Family Safety
\item MMGuardian Parent App
\item MMGuardian Parental Control
\item Mobicip Parental Controls
\item MobiLinc Cam Viewer
\item MovieStarPlanet
\item mSpy Lite - Phone tracker app
\item MyHeritage: Family Tree \& DNA
\item Net Nanny Parental Control App
\item Norton Family Parental Control
\item Off Remote
\item Offender Locator
\item Offender Locator Lite
\item Online Walkie Talkie Pro
\item OurFamilyWizard Co-Parent App
\item OurPact Jr. Child App
\item Parental Control \& Screen Time
\item Parental Control App - unGlue
\item Parental Control Kid’s App
\item Parental Control Parent’s App
\item ParentKit - Parental Controls for iOS
\item People Tracker Pro - Cell Phone Tracker App!
\item Periscope Live Video Streaming
\item Phone Tracker By Number
\item Phone Tracker for iPhones
\item Phone Tracker for iPhones (Track people with GPS)
\item Pingo by Findmykids
\item Placeter
\item PocketFinder 2
\item Presence: Video Security
\item Qustodio Parental Control
\item Qustodio Parental Control App
\item React Mobile – Safety App
\item Safe Family: Screen Time App
\item Safe SMS
\item SaferKid Text Monitoring App
\item Safety App for Silent Beacon
\item Screen Time Parental Control
\item ScreenGuide Parental Control
\item SecureTeen Parental Control
\item Securly Home
\item SeTracker2
\item Share Location: Phone Tracker
\item Skout — Meet New People
\item Smart Family Companion
\item SoSecure by ADT: Safety App
\item Spoten Phone Location Tracker
\item StudentVUE
\item T-Mobile FamilyMode
\item Tango - Video Call \& Chat
\item TeenOrbit Parent Control Panel
\item Text Free: Texting App + SMS
\item Text Me - Phone Call + Texting
\item TextFree: Call + Texting Line
\item TextNow
\item Two Way : Walkie Talkie
\item unGlue Kids
\item Verizon FamilyBase
\item Viber: Free Calling \& Texting
\item Voxer Walkie Talkie Messenger
\item WebWatcher Parent App
\item WeSeeYou Safety App
\item WhereAreYou App Locate friends
\item Whisper - Share, Express, Meet
\item Wink - Dating \& Friends
\item Wizz App - chat now
\item XFINITY xFi
\item Yik Yak
\item Zenly - your social map
\end{inparaitem}

\end{document}